\newif\ifdraft
\newif\iffull
\newif\ifcomment
\newif\iflatexdiff
\newif\ifbibtex
\newif\ifpreprint
\preprinttrue    
\fulltrue        
\newif\ifplbpaper
\newif\ifapspaper
\apspapertrue
\def\snntitle{$\snn$}
\ifpreprint
\def\snntitle{$\snnbf$}
\fi
\def\dvers{v1.00}       
\def\dtitle{Centrality determination of \PbPb\ collisions \\at \snntitle = 2.76 TeV with ALICE}  
\def\stitle{Centrality determination with ALICE} 
\def\phnum{2012-368}     
\def\phdat{18 Dec 2012}  
\def\bibstname{apsrev4-1}
\ifpreprint
\documentclass[ALICE,manyauthors,12pt]{cernphprep}
\usepackage[comma,square,numbers,sort&compress]{natbib}
\else 
\ifplbpaper
\documentclass[final,3p,12pt]{elsarticle}
\biboptions{comma,square,numbers,sort&compress}
\def\bibstname{utphys}
\fi
\ifapspaper
\documentclass[10pt,aps,superscriptaddress,altaffilletter,nobibnotes,nofootinbib]{revtex4-1}
\newenvironment{frontmatter}{}{\maketitle}
\def\bibstname{apsrev4-1}
\fi
\fi
\usepackage{graphicx}  
\usepackage{dcolumn}   
\usepackage{bm}        
\usepackage{amssymb}   
\usepackage{amsfonts}
\usepackage{graphics}
\usepackage{grffile}   
\usepackage{epsfig}
\usepackage{units}
\usepackage{hyperref}
\usepackage[usenames]{color}
\usepackage[normalem]{ulem} 
\usepackage{color}
\usepackage[utf8]{inputenc}
\usepackage[T1]{fontenc}
\iflatexdiff
\RequirePackage{color}\definecolor{RED}{rgb}{1,0,0}\definecolor{BLUE}{rgb}{0,0,1}

\fi
\newcommand{\ITS}          {\rm{ITS}}

\newcommand{\ZEM}          {\rm{ZEM}}
\newcommand{\ZDC}          {\rm{ZDC}}
\newcommand{\SPD}          {\rm{SPD}}

\newcommand{\TPC}          {\rm{TPC}}

\newcommand{\VZERO}        {\rm{VZERO}}
\newcommand{\VZEROA}       {\rm{VZERO-A}}
\newcommand{\VZEROC}       {\rm{VZERO-C}}

\newcommand{\pp}           {pp}
\newcommand{\ppbar}        {\mbox{$\mathrm {p\overline{p}}$}}
\newcommand{\PbPb}         {\mbox{Pb--Pb}}

\newcommand{\pA}           {\mbox{p--A}}
\newcommand{\dA}           {\mbox{d--A}}
\newcommand{\AAA}          {\mbox{A--A}}

\newcommand{\Nch}          {\ensuremath{N_\mathrm{ch}}}
\newcommand{\dNdeta}       {\mathrm{d}N_\mathrm{ch}/\mathrm{d}\eta}

\newcommand{\snn}          {\ensuremath{\sqrt{s_{\rm NN}}}}
\newcommand{\snnbf}        {\ensuremath{\mathbf{{\sqrt{s_{\mathbf NN}}}}}}
\newcommand{\sonly}        {\ensuremath{\sqrt{s}}}
\newcommand{\Nanc}         {\ensuremath{N_\mathrm{ancestors}}}
\newcommand{\TAB}          {\ensuremath{T_\mathrm{AA}}}
\newcommand{\Nspec}        {\ensuremath{N_\mathrm{spec}}}
\newcommand{\Npart}        {\ensuremath{N_\mathrm{part}}}
\newcommand{\avNpart}      {\ensuremath{\langle N_\mathrm{part} \rangle}}
\newcommand{\avNpartdata}  {\ensuremath{\langle N_\mathrm{part}^{\rm data} \rangle}}
\newcommand{\avNpartgeo}   {\ensuremath{\langle N_\mathrm{part}^{\rm geo} \rangle}}
\newcommand{\Ncoll}        {\ensuremath{N_\mathrm{coll}}}
\newcommand{\avNcoll}      {\ensuremath{\langle N_\mathrm{coll} \rangle}}

\newcommand{\avTAB}        {\ensuremath{\langle T_\mathrm{AA} \rangle}}

\newcommand{\dNdetapt}     {\ensuremath{\dNdeta\,/\left(0.5\Npart\right)}}

\newcommand{\signn}        {\ensuremath{\sigma^{\rm inel}_{\rm NN}}}

\newcommand{\syst}         {({\it sys.})}
\newcommand{\Fig}[1]       {Fig.~\ref{#1}}
\newcommand{\Figure}[1]    {Figure~\ref{#1}}

\graphicspath{{./img/}}
\ifdraft
\usepackage{lineno}
\linenumbers
\modulolinenumbers[5]
\usepackage{fancyhdr}
\fi
\begin{document}
\ifpreprint
\begin{titlepage}
\PHnumber{\phnum}    
\PHdate{\phdat}
\title{\dtitle}
\ShortTitle{\stitle}
\iffull
\Collaboration{ALICE Collaboration~\thanks{See Appendix~\ref{app:collab} for the list of collaboration members}}
\else
\Collaboration{ALICE Collaboration}
\fi
\ShortAuthor{ALICE Collaboration} 
\ifdraft
\begin{center}
\today\\ \color{red}DRAFT \dvers\ \hspace{0.3cm} \$Revision: 154 $\color{white}:$\$\color{black}\vspace{0.3cm}
\end{center}
\fi
\else
\begin{frontmatter}
\title{\dtitle}
\iffull
\input{authors-paper.tex}            
\else
\ifdraft
\author{ALICE Collaboration \\ \vspace{0.3cm} 
\today\\ \color{red}DRAFT \dvers\ \hspace{0.3cm} \$Revision: 154 $\color{white}:$\$\color{black}}
\else
\author{ALICE Collaboration}
\fi
\fi
\fi
\begin{abstract}
  This publication describes the methods used to measure the centrality
  of inelastic Pb--Pb collisions at a center-of-mass energy of 2.76 TeV 
  per colliding nucleon pair with ALICE. 
  The centrality is a key parameter in the study of the properties of 
  QCD matter at extreme temperature and energy density, because it is
  directly related to the initial overlap region of the colliding nuclei.
  Geometrical properties of the collision, such as the number of
  participating nucleons and number of binary nucleon-nucleon
  collisions, are deduced from a Glauber model with a sharp impact
  parameter selection, and shown to be consistent with those extracted
  from the data.
  The centrality determination provides a tool to compare ALICE measurements 
  with those of other experiments and with theoretical calculations.

\ifdraft 
\ifpreprint
\end{abstract}
\end{titlepage}
\else
\end{abstract}
\end{frontmatter}
\newpage
\fi
\fi
\ifdraft
\thispagestyle{fancyplain}
\else
\end{abstract}
\ifpreprint
\end{titlepage}
\else
\end{frontmatter}
\fi
\fi
\setcounter{page}{2}

\section{Introduction}
\label{sec:introduction}

Ultra-relativistic heavy-ion collisions at the Large Hadron Collider~(LHC) 
produce strongly interacting matter under extreme conditions of temperature
and energy density, similar to those prevailing in the first few
microseconds after the Big Bang \cite{bbang}.

Since nuclei are extended objects, the volume of the interacting region
depends on the impact parameter ($b$) of the collision, defined as the
distance between the centers of the two colliding nuclei in a plane
transverse to the beam axis.
It is customary in the field of heavy-ion physics to introduce the concept of 
the centrality of the collision, which is directly related to the impact
parameter, and inferred by comparison of data with simulations of the collisions.

The purely geometrical Glauber model~\cite{Miller:2007ri}, which typically
is used in this context, has its origins in the quantum mechanical model for
\pA\ and \AAA\ scattering described in ~\cite{glauber1955, glauber1959, 
glauber2006}. The model treats a nuclear collision as
a superposition of binary nucleon-nucleon interactions. The volume of
the initial overlap region is expressed via the number of participant
nucleons.  A participant nucleon of one nucleus is defined as a
nucleon that undergoes one or more binary collisions with nucleons of
the other nucleus.  The number of participants and spectators, \Npart\
and \mbox{$\Nspec=2A-\Npart$}, where A is the total number of nucleons
in the nucleus (mass number), and the number of binary collisions
\Ncoll\ are calculated for a given value of the impact parameter and
for a realistic initial distribution of nucleons inside the nucleus,
and assuming that nucleons follow straight trajectories.  This
approach provides a consistent description of \pA, \dA, and \AAA\
collisions, and is especially useful when comparing data from
different experiments or from different collision systems and to
theoretical calculations.

Neither the impact parameter nor geometrical quantities, such as
\Npart, \Nspec, or \Ncoll\ are directly measurable. Two experimental
observables related to the collision geometry are the average
charged-particle multiplicity \Nch\ and the energy carried by
particles close to the beam direction and deposited in Zero-Degree
Calorimeters (\ZDC), called the zero-degree energy $E_{ZDC}$. The
average charged-particle multiplicity is assumed to decrease
monotonically with increasing impact parameter.  The energy deposited
in the zero-degree calorimeters, $E_{ZDC}$, is directly related to the
number of spectator nucleons \Nspec, which constitute the part of the
nuclear volume not involved in the interaction. However, unlike \Nch,
$E_{ZDC}$ does not depend monotonically on the impact parameter $b$
because nucleons bound in nuclear fragments with similar magnetic
rigidity as the beam nuclei remain inside the beam-pipe and therefore
are not detected in the ZDC. Since fragment formation is more
important in peripheral collisions, the monotonic relationship between
$E_{ZDC}$ and $b$ is valid only for relatively central events (small
$b$).  For this reason, the zero-degree energy measurement needs to be
combined with another observable that is monotonically correlated with $b$.

The centrality is usually expressed as a percentage of the total
nuclear interaction cross section $\sigma$~\cite{Miller:2007ri}.
The centrality percentile $c$ of an \AAA\ collision with an impact parameter 
$b$ is defined by integrating the impact parameter distribution $\mathrm{d}\sigma/
\mathrm{d}b^{'}$ as
\begin{equation}
c =\frac{\int_0^{b}{\mathrm{d}\sigma / \mathrm{d}b^{\prime}} \, \mathrm{d}b^{\prime}}
    {\int_0^{\infty}{\mathrm{d}\sigma / \mathrm{d}b^{\prime}} \, \mathrm{d}b^{\prime}} 
  = \frac{1}{\sigma_{AA}} \, \int_0^{b}{\frac{\mathrm{d}\sigma}{\mathrm{d}b^{\prime}}} \, \mathrm{d}b^{\prime}.
\end{equation}

In ALICE, the centrality is defined as the percentile of the hadronic
cross section corresponding to a particle multiplicity above a given
threshold ($N_{ch}^{THR}$) or an energy deposited in the \ZDC\
below a given value ($E_{ZDC}^{THR}$) in the \ZDC\ energy distribution $\mathrm{d}\sigma/\mathrm{d}E_{ZDC}^{\prime}$
\begin{equation}
c \approx \frac{1}{\sigma_{AA}} \, \int_{N_{ch}^{THR}}^{\infty}{\frac{\mathrm{d}\sigma}{\mathrm{d}\Nch^{\prime}}} \, \mathrm{d}\Nch^{\prime}
  \approx \frac{1}{\sigma_{AA}} \, \int_{0}^{E_{ZDC}^{THR}}{\frac{\mathrm{d}\sigma}{\mathrm{d}E_{ZDC}^{\prime}}} \, \mathrm{d}E_{ZDC}^{\prime}. 
\end{equation}
The procedure can be simplified by replacing the cross section with
the number of observed events, corrected for the trigger efficiency.
However, at LHC energies, the strong electromagnetic fields generated
by the heavy ions moving at relativistic velocity lead to large cross
sections for QED processes
\cite{Bruce2009,Djuvsland:2010qs,HEN03-QED,Oppedisano:2011}.  Although
the cross sections for these processes exceed those for the hadronic cross
section by several orders of magnitude, they only contaminate the
hadronic cross section in the most peripheral collisions.  For this
reason one may choose to restrict the centrality determination to the
region where such contamination is negligible.  The fraction of
hadronic events excluded by such cut as well as the trigger efficiency
can be estimated using a model of the nuclear collision and the
related particle production.

In this paper, we report on the centrality determination used in the analyses 
of the \PbPb\ collision data from the 2010 and 2011 run recorded
with the ALICE detector~\cite{aliceapp}.
Specifically the analysis presented here is done with a subset of the 2010 data, 
but the methods and results are valid for 2011 as well. 
In Section~\ref{sec:glauber}, we describe the implementation of the
Glauber model used by ALICE.  We extract mean numbers of the relevant
geometrical quantities for typical centrality classes defined by
classifying the events according to their impact parameter.
Section~\ref{sec:alice} describes the experimental conditions and the
event selection with particular emphasis on the rejection of QED and
machine-induced backgrounds.  
Section~\ref{sec:mult} presents the methods employed by ALICE for the
determination of the hadronic cross section, needed for the absolute
determination of the centrality. The main method uses the \VZERO\
amplitude distribution fitted with the Glauber model. The systematic
uncertainty is obtained by comparing the fit to an unbiased \VZERO\
distribution obtained by correcting the measured one by the efficiency
of the event selection and the purity of the event sample.
Section \ref{sec:zdc} presents the determination of the centrality
classes using either the multiplicity at mid-rapidity or the energy
deposited in the \ZDC. We discuss the relation between the measured
multiplicity and geometrical quantities connected to centrality,
established by the Glauber Model. These are nearly identical to those
obtained in Section~\ref{sec:glauber}, classifying the events
according to their impact parameter, which are therefore used as
reference in all ALICE analyses.
Section~\ref{sec:resolution} presents the precision of the
centrality determination in ALICE.
Section~\ref{sec:conclusions} summarizes and concludes the paper.

\section{The Glauber Model}
\label{sec:glauber}

\begin{figure}[t!f]
 \centering
 \includegraphics[width=0.75\textwidth]{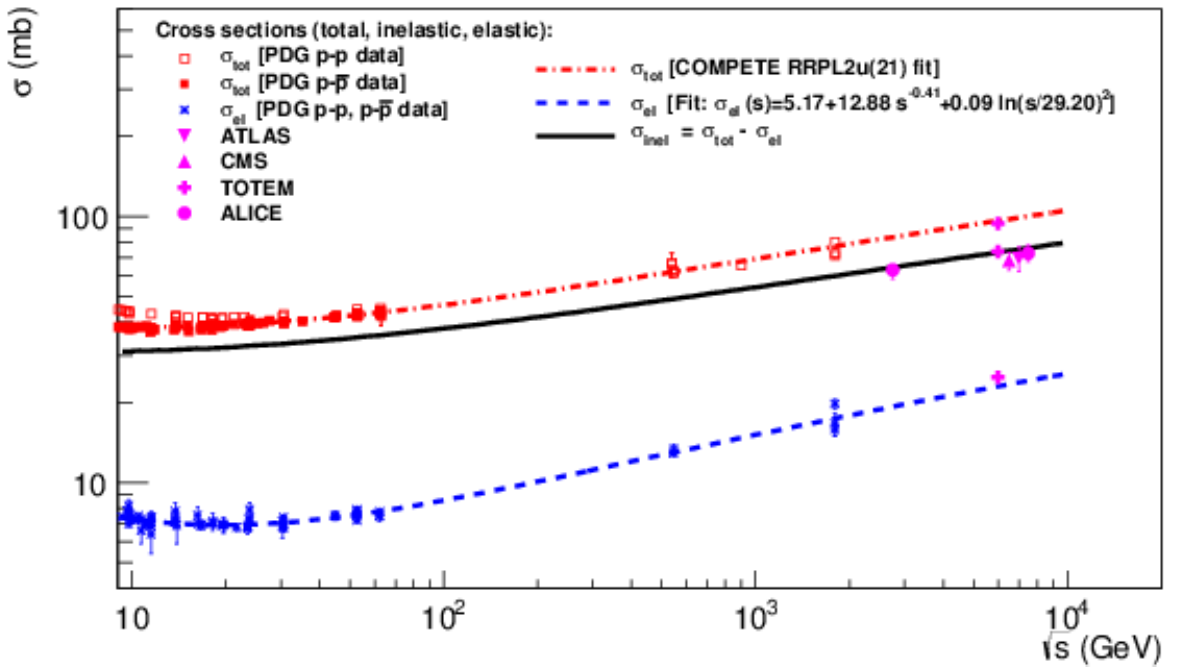}
 \caption{
  (Color online) Compilation of total $\sigma_{\rm NN}^{tot}$, elastic
  $\sigma_{\rm NN}^{el}$, and inelastic $\sigma_{\rm NN}^{inel}$ cross
  sections of pp and p$\bar{\mathrm {p}}$ collisions
  \cite{Denterria2011, Nakamura:2010zzi, Block}. The $\sigma_{\rm NN}^{el}$ curve is a fit
  performed by the COMPETE Collaboration also available at
  \cite{Nakamura:2010zzi, COMPETE}. The \pp\ data from ATLAS
  \cite{ATLAScrossec}, CMS \cite{CMScrossec}, TOTEM,
  \cite{TOTEMcrossec} and ALICE \cite{Poghosyan:2011} agree well with
  the interpolation for \signn.
  \label{fig:sigmaNN}}
\end{figure}

The Glauber model is widely used to describe the dependence of \Npart\
and \Ncoll\ on $b$ in \pA, \dA\ and \AAA\ collisions
\cite{Miller:2007ri, glauber1955, glauber1959, glauber2006}.  The
purpose of Monte Carlo implementations of the Glauber model
\cite{glauberMC, glauberMC2} is to compose two nuclei out of nucleons
and simulate their collision process event-by-event.  Geometrical
quantities are calculated by simulating many nucleus-nucleus
collisions.  Mean values of these quantities are calculated for
centrality classes defined by classifying the events according to
their impact parameter $b$.

Following \cite{Alver:2008aq}, the first step in the Glauber Monte Carlo is to
prepare a model of the two nuclei by defining stochastically the
position of the nucleons in each nucleus. The nucleon position in the
$^{208}$Pb nucleus is determined by the nuclear density function,
modeled by the functional form (modified Woods-Saxon or 2-parameter
Fermi distribution):
\begin{equation}
\rho(r) = \rho_{0} \frac{1+w(r/R)^2}{1 + \exp \left(\frac{r-R}{a}\right) }
\end{equation}
The parameters are based on data from low energy electron-nucleus
scattering experiments~\cite{DeJager:1987qc}.  Protons and neutrons
are assumed to have the same nuclear profile.  The parameter
$\rho_{0}$ is the nucleon density, which provides the overall
normalization, not relevant for the Monte Carlo simulation, $R=(6.62
\pm 0.06)$~fm is the radius parameter of the $^{208}$Pb nucleus and
$a=(0.546 \pm 0.010)$~fm is the skin thickness of the nucleus, which
indicates how quickly the nuclear density falls off near the edge of
the nucleus.  The additional parameter $w$ is needed to describe
nuclei whose maximum density is reached at radii $r>0$ ($w=0$ for Pb).
In the Monte Carlo procedure the radial coordinate of a nucleon is
randomly drawn from the distribution $4 \pi r^2 \rho(r)$ and $\rho_0$
is determined by the overall normalization condition $\int \rho(r)
d^3r = A$.  We require a hard-sphere exclusion distance of
$d_\mathrm{min}=0.4$~fm between the centers of the nucleons, i.e. no
pair of nucleons inside the nucleus has a distance less than
$d_\mathrm{min}$. The hard-sphere exclusion distance, characteristic
of the length of the repulsive nucleon-nucleon force, is not known
experimentally and thus is varied by 100\% ($d_\mathrm{min} = (0.4 \pm
0.4)$~fm).

The second step is to simulate a nuclear collision. The impact
parameter $b$ is randomly selected from the geometrical distribution
$dP/db \sim b$ up to a maximum $b_\mathrm{max}\simeq 20 \,
\mathrm{fm}>2R_{\rm Pb}$. The maximum value of the impact parameter
$b_\mathrm{max}$ is chosen large enough to simulate collisions until
the interaction probability becomes zero. This is particularly
important for the calculation of the total \PbPb\ cross section.
The nucleus-nucleus collision is treated as a sequence of independent
binary nucleon-nucleon collisions, where the nucleons travel on
straight-line trajectories and the inelastic nucleon-nucleon
cross section is assumed to be independent of the number of collisions
a nucleon underwent previously, i.e.\ the same cross section is used
for all successive collisions.
Two nucleons from different nuclei are assumed to collide if the
relative transverse distance between centers is less than the distance
corresponding to the inelastic nucleon-nucleon cross section
$d<\sqrt{\signn/\pi}$.  A Gaussian overlap function can be used as an
alternative to the black-disk nucleon-nucleon overlap function
\cite{Rybczynski:2011wv}. It makes no significant difference within systematic
uncertainty in the global event properties.

\begin{table*}[thb!f]
 \footnotesize 
 \centering
 \caption{Geometric properties (\Npart, \Ncoll, \TAB) of \PbPb\ collisions for centrality classes defined by 
         sharp cuts in the impact parameter $b$ (in fm). The mean values, the RMS, and the systematic uncertainties 
         are obtained with a Glauber Monte Carlo calculation. 
 \label{tab:Npart}}
 \hspace{-1cm}\begin{tabular}{c|ccccccccccc}
   Centrality & $b_{\rm min}$ & $b_{\rm max}$ & $\langle\Npart\rangle$ & RMS & $\syst$ & $\langle\Ncoll\rangle$ & RMS & $\syst$ & $\langle\TAB\rangle$ & RMS & $\syst$ \\
        & (fm) & (fm) & & & & & & & 1/mbarn & 1/mbarn & 1/mbarn \\
\hline
0--1\%    & 0.00   & 1.57  & 403.8 & 4.9 & 1.8  & 1861  & 82  & 210   & 29.08   & 1.3   & 0.95   \\
1--2\%    & 1.57   & 2.22  & 393.6 & 6.5 & 2.6  & 1766  & 79  & 200   & 27.6    & 1.2   & 0.87   \\
2--3\%    & 2.22   & 2.71  & 382.9 & 7.7 & 3.0  & 1678  & 75  & 190   & 26.22   & 1.2   & 0.83   \\
3--4\%    & 2.71   & 3.13  & 372.0 & 8.6 & 3.5  & 1597  & 72  & 180   & 24.95   & 1.1   & 0.81   \\
4--5\%    & 3.13   & 3.50  & 361.1 & 9.3 & 3.8  & 1520  & 70  & 170   & 23.75   & 1.1   & 0.81   \\
5--10\%   & 3.50   & 4.94  & 329.4 & 18  & 4.3  & 1316  & 110 & 140   & 20.56   & 1.7   & 0.67   \\
10--15\%  & 4.94   & 6.05  & 281.2 & 17  & 4.1  & 1032  & 91  & 110   & 16.13   & 1.4   & 0.52   \\
15--20\%  & 6.05   & 6.98  & 239.0 & 16  & 3.5  & 809.8 & 79  & 82    & 12.65   & 1.2   & 0.39   \\
20--25\%  & 6.98   & 7.81  & 202.1 & 16  & 3.3  & 629.6 & 69  & 62    & 9.837   & 1.1   & 0.30   \\
25--30\%  & 7.81   & 8.55  & 169.5 & 15  & 3.3  & 483.7 & 61  & 47    & 7.558   & 0.96  & 0.25   \\
30--35\%  & 8.55   & 9.23  & 141.0 & 14  & 3.1  & 366.7 & 54  & 35    & 5.73    & 0.85  & 0.20   \\
35--40\%  & 9.23   & 9.88  & 116.0 & 14  & 2.8  & 273.4 & 48  & 26    & 4.272   & 0.74  & 0.17   \\
40--45\%  & 9.88   & 10.47 & 94.11 & 13  & 2.6  & 199.4 & 41  & 19    & 3.115   & 0.64  & 0.14   \\
45--50\%  & 10.47  & 11.04 & 75.3  & 13  & 2.3  & 143.1 & 34  & 13    & 2.235   & 0.54  & 0.11   \\
50--55\%  & 11.04  & 11.58 & 59.24 & 12  & 1.8  & 100.1 & 28  & 8.6   & 1.564   & 0.45  & 0.082  \\
55--60\%  & 11.58  & 12.09 & 45.58 & 11  & 1.4  & 68.46 & 23  & 5.3   & 1.07    & 0.36  & 0.060  \\
60--65\%  & 12.09  & 12.58 & 34.33 & 10  & 1.1  & 45.79 & 18  & 3.5   & 0.7154  & 0.28  & 0.042  \\
65--70\%  & 12.58  & 13.05 & 25.21 & 9.0 & 0.87 & 29.92 & 14  & 2.2   & 0.4674  & 0.22  & 0.031  \\
70--75\%  & 13.05  & 13.52 & 17.96 & 7.8 & 0.66 & 19.08 & 11  & 1.3   & 0.2981  & 0.17  & 0.020  \\
75--80\%  & 13.52  & 13.97 & 12.58 & 6.5 & 0.45 & 12.07 & 7.8 & 0.77  & 0.1885  & 0.12  & 0.013  \\
80--85\%  & 13.97  & 14.43 & 8.812 & 5.2 & 0.26 & 7.682 & 5.7 & 0.41  & 0.12    & 0.089 & 0.0088 \\
85--90\%  & 14.43  & 14.96 & 6.158 & 3.9 & 0.19 & 4.904 & 4.0 & 0.24  & 0.07662 & 0.062 & 0.0064 \\
90--95\%  & 14.96  & 15.67 & 4.376 & 2.8 & 0.10 & 3.181 & 2.7 & 0.13  & 0.0497  & 0.042 & 0.0042 \\
95--100\% & 15.67  & 20.00 & 3.064 & 1.8 & 0.059& 1.994 & 1.7 & 0.065 & 0.03115 & 0.026 & 0.0027 \\

\hline
\hline
 0--5\%    & 0.00 & 3.50   & 382.7 & 17  & 3.0  & 1685  & 140 & 190  & 26.32   & 2.2   & 0.85\\
 5--10\%   & 3.50 & 4.94   & 329.4 & 18  & 4.3  & 1316  & 110 & 140  & 20.56   & 1.7   & 0.67\\
 10--20\%  & 4.94 & 6.98   & 260.1 & 27  & 3.8  & 921.2 & 140 & 96   & 14.39   & 2.2   & 0.45\\
 20--40\%  & 6.98 & 9.88   & 157.2 & 35  & 3.1  & 438.4 & 150 & 42   & 6.850   & 2.3   & 0.23\\
 40--60\%  & 9.88 & 12.09  & 68.56 & 22  & 2.0  & 127.7 & 59  & 11   & 1.996   & 0.92  & 0.097\\
 60--80\%  & 12.09& 13.97  & 22.52 & 12  & 0.77 & 26.71 & 18  & 2.0  & 0.4174  & 0.29  & 0.026\\
 80--100\% & 13.97& 20.00  & 5.604 & 4.2 & 0.14 & 4.441 & 4.4 & 0.21 & 0.06939 & 0.068 & 0.0055\\
\end{tabular}
\end{table*}

The number of collisions \Ncoll\ and the number of participants
\Npart\ are determined by counting, respectively, the binary nucleon
collisions and the nucleons that experience at least one collision.
Following the notation in \cite{Miller:2007ri}, the geometric nuclear
overlap function \TAB\ is then calculated as $\TAB = \Ncoll /
\signn$, and represents the effective nucleon luminosity in the 
collision process.

For nuclear collisions at \snn~$ = 2.76$~TeV, we use
\signn\ = $(64~\pm 5)$~mb, estimated by
interpolation~\cite{Denterria2011} of \pp\ data at different
center-of-mass energies and from cosmic rays~\cite{Nakamura:2010zzi,
  COMPETE}, and subtracting the elastic scattering cross section from
the total cross section.  The interpolation is in good agreement with
the ALICE measurement of the \pp\ inelastic cross section at \snn~$ =
2.76$~TeV, \signn\ = $(62.8~\pm 2.4 ^{+1.2}_{-4.0})$~mb
\cite{Poghosyan:2011}, and with the measurements of ATLAS
\cite{ATLAScrossec}, CMS \cite{CMScrossec}, and TOTEM \cite{TOTEMcrossec}
 at \snn =~7~TeV, as shown in
Fig.~\ref{fig:sigmaNN}.

The total \PbPb\ cross section is calculated as
$\sigma_{\mathrm{PbPb}} = N_{\rm evt}(\Ncoll \geq 1)/N_{\rm evt}(\Ncoll \geq 0) 
\times \pi b_{\mathrm{max}}^2$, i.e.\ the geometrical value corrected by
the fraction of events with at least one nucleon-nucleon collision.
We obtain $\sigma_{\mathrm{PbPb}} = (7.64 \pm 0.22 (syst.))$~b, in agreement with the ALICE 
measurement $\sigma_{\mathrm{PbPb}} = (7.7 \pm 0.1 (stat.) ^{+0.6}_{-0.5} (syst.))$~b~\cite{Oppedisano:2011}.

\begin{figure}[tbh!f]
 \centering
 \includegraphics[width=0.49\textwidth]{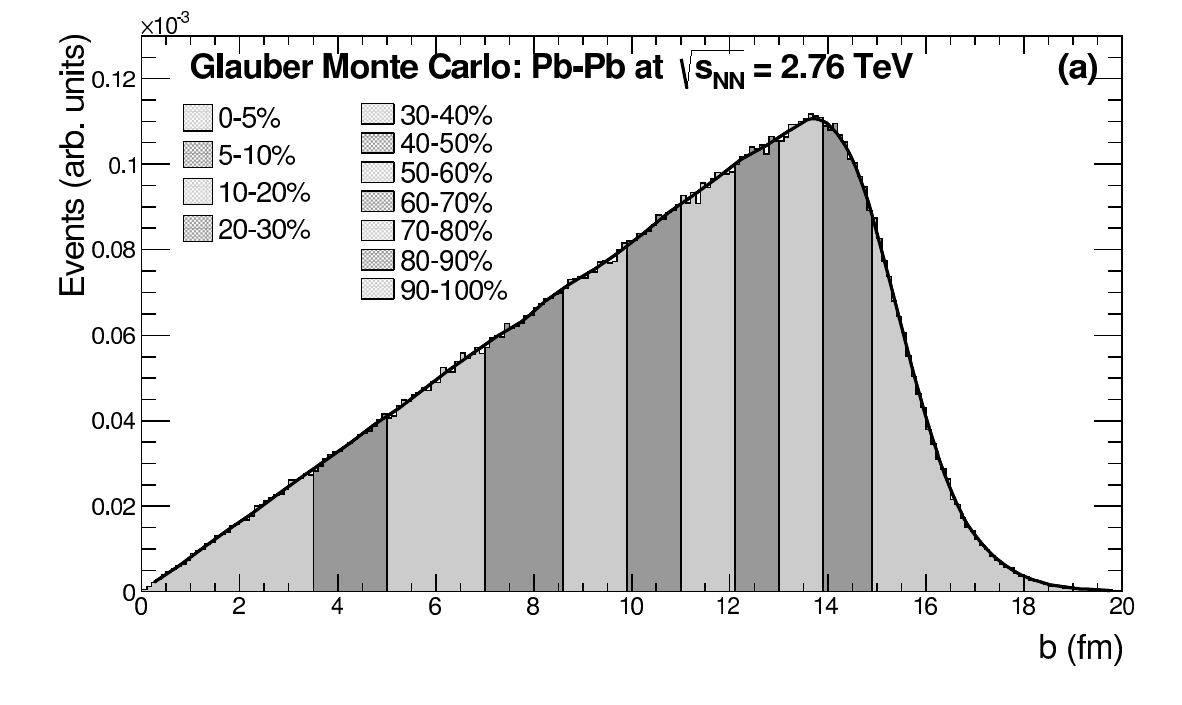}
 \includegraphics[width=0.49\textwidth]{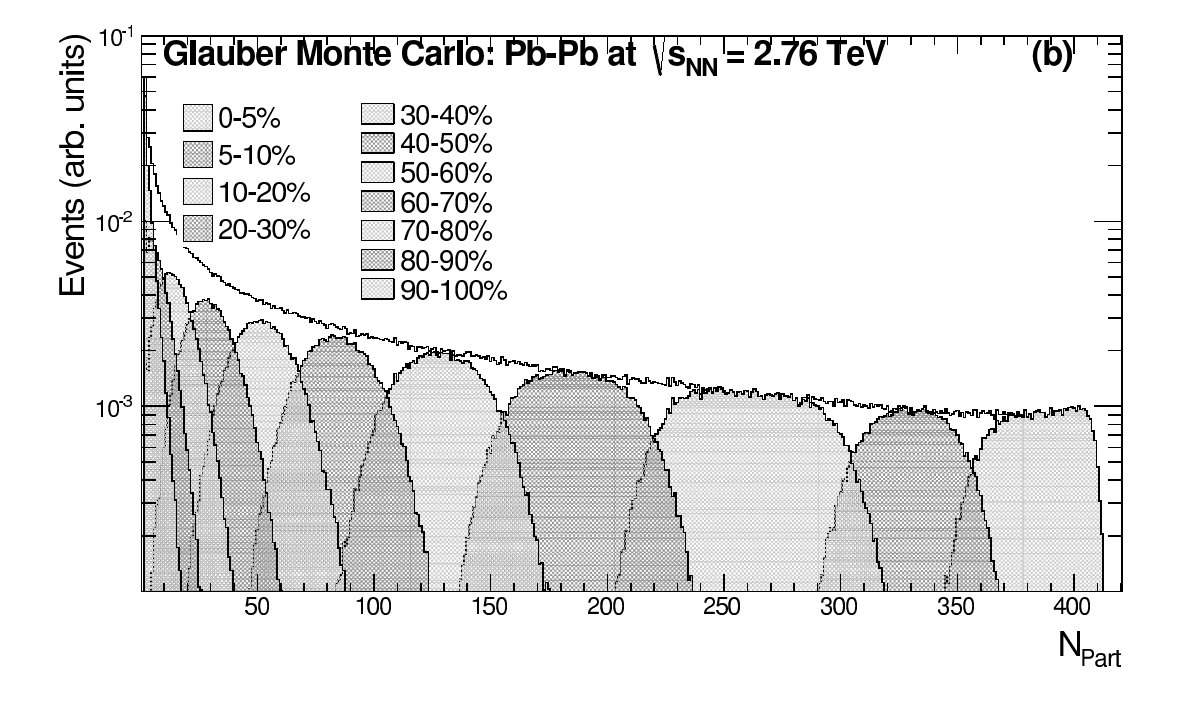}
 \caption{Geometric properties of \PbPb\ collisions at
  $\snn=2.76$~TeV obtained from a Glauber Monte Carlo calculation:
  Impact parameter distribution (left), sliced for percentiles of the
  hadronic cross section, and distributions of the number of
  participants (right) for the corresponding centrality classes.
  \label{fig:glaugeo}}
\end{figure}

Table~\ref{tab:Npart} report the mean number of participants \avNpart\
and collisions \avNcoll, and the mean nuclear thickness function
\avTAB\ for centrality classes defined by sharp cuts in the impact
parameter $b$ calculated with the Glauber model~(Fig.~\ref{fig:glaugeo}).
The root mean square (RMS) of these distributions 
is a measure for the magnitude of the dispersion of the quantities.

\begin{figure}[tbh!f]
 \centering
 \includegraphics[width=0.49\textwidth]{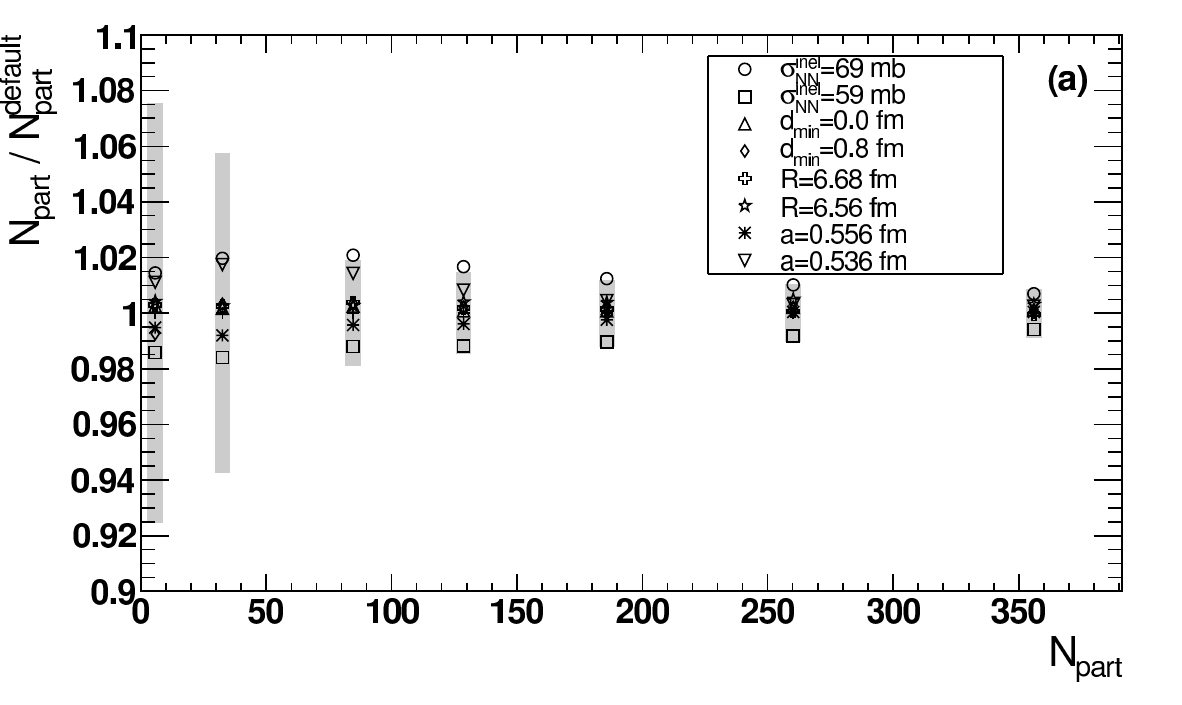}
 \includegraphics[width=0.49\textwidth]{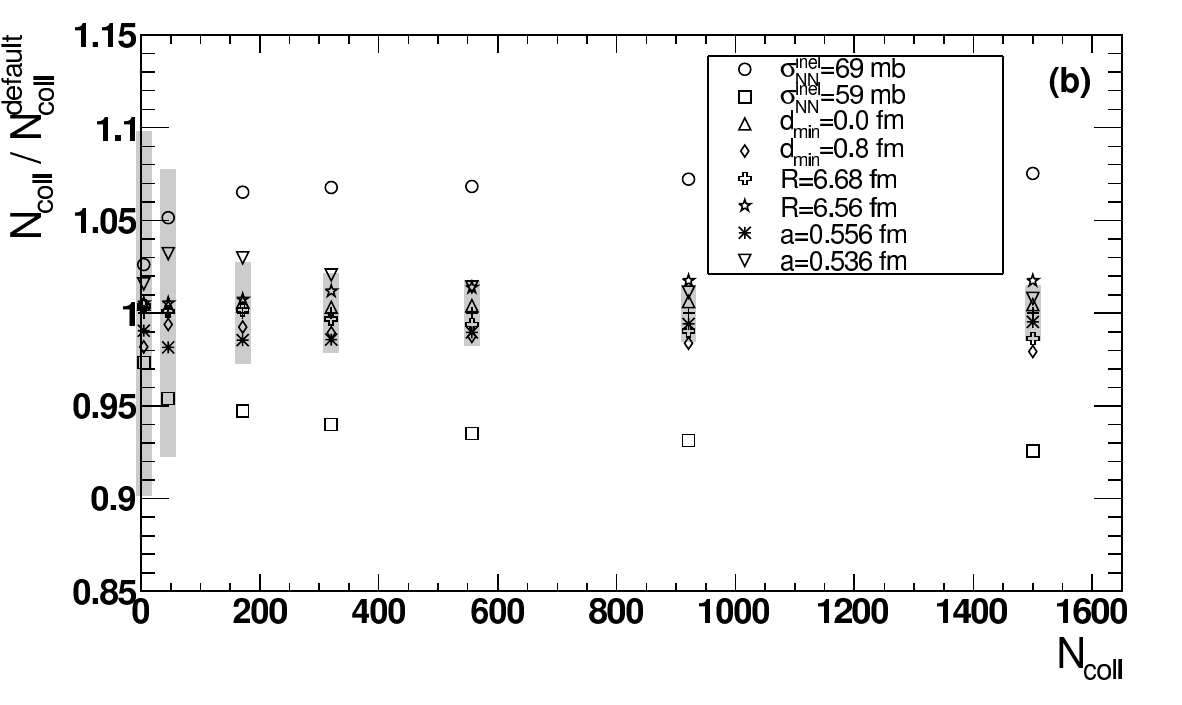}
 \caption{Sensitivity of \Npart\ (left) and \Ncoll\ (right) to
  variations of parameters in the Glauber Monte Carlo model of
  \PbPb\ collisions at \snn = 2.76 TeV. The gray band represents the
  RMS of \Npart\ and \Ncoll\ respectively. It is scaled by a factor 0.1 for visibility.
  \label{fig:glausys}}
\end{figure}

The systematic uncertainties on the mean values are obtained by
independently varying the parameters of the Glauber model within their
estimated uncertainties. More specifically, the default value of the
nucleon--nucleon cross section of \signn\ = 64 mb was varied between 
59 mb and 69 mb. The Woods-Saxon parameters were varied by one standard 
deviation to determine uncertainties related to the nuclear density profile.  
The minimum distance of 0.4~fm between two nucleons of the same nucleus was 
varied by 100\%, from 0 to 0.8 fm to evaluate the effects of a nucleon hard
core~(as mentioned above). Figure~\ref{fig:glausys} shows the resulting 
variations for \PbPb\ collisions at \snn = 2.76 TeV.
The total systematic uncertainty reported in Table~\ref{tab:Npart}
was obtained by adding in quadrature the deviations from the default
result for each of the variations listed above. The uncertainty of
\Npart\ ranges from about 3--4\% in peripheral collisions to $<$ 1\% in
central collisions, the uncertainty of \Ncoll\ ranges from about 7\%
in peripheral collisions to about 11\% in central collisions, the
uncertainty of \TAB\ ranges from about 6\% in peripheral collisions to
about 3\% in central collisions.  The nuclear overlap function \TAB\
is often used to compare observables related to hard processes in
\AAA\ and \pp\ collisions. Since $\TAB = \Ncoll / \signn$, it has the
same systematic uncertainties as \Ncoll\, except that the uncertainty
on \signn\ cancels out.

Finally it is worth noting that more sophisticated implementations of the Glauber 
model~\cite{Broniowski:2007ft,Broniowski:2010jd,Rybczynski:2011wv} suggest that 
effects not included in our Glauber model, such as the changes of the excluded 
volume on the nuclear density and two-body correlations, can be approximated 
by slightly adjusting the Woods-Saxon parameters. The modified parameters, however, 
are well covered by the systematic uncertainty quoted above for the parameters 
that we use.


\section{Experimental Conditions}
\label{sec:alice}

\subsection{The ALICE detector}
\label{subsec:alice}

ALICE is an experiment dedicated to the study of heavy-ion
collisions at the LHC.  A detailed description of the apparatus is given
in Ref.~\cite{aliceapp}.  Here, we briefly describe the detector
components used in this analysis.

The Silicon Pixel Detector (\SPD) is the innermost part of the Inner
Tracking System~(\ITS).  It consists of two cylindrical layers of
hybrid silicon pixel assemblies positioned at average radial distances of
$3.9$ and $7.6$~cm from the beam line, with a total of $9.8\times
10^{6}$ pixels of size $50\times425$~$\mu$m$^2$, read out by 1200
electronic chips.  The \SPD\ coverage for particles originating from
the center of the detector is $|\eta|<2.0$ and $|\eta|<1.4$ for the
inner and outer layers, respectively.  Each chip provides a fast
signal if at least one of its pixels is hit.  The signals from the
1200 chips are combined in a programmable logic unit which supplies a
trigger signal.  The fraction of \SPD\ channels active during 2010 data
taking was 70\% for the inner and 78\% for the outer layers.

The \VZERO\ detector consists of two arrays of 32 scintillator cells
placed at distances $z=3.4$~m and $z=-0.9$~m from the nominal
interaction point, along the beam line, covering the full azimuth. 
The \VZERO\ detector is within $2.8<\eta<5.1$~(\VZEROA) and $-3.7<\eta<-1.7$~(\VZEROC).  
Both amplitude and time of signals in each scintillator are recorded.  The
\VZERO\ time resolution is better than 1~ns, allowing discrimination
of beam--beam collisions from background events produced upstream of
the experiment.  The \VZERO\ is also used to provide a trigger signal
(see \ref{subsec:event-selection}).

The Time Projection Chamber (\TPC) is used for charged particle
trajectory reconstruction, track momentum measurement and particle
identification. The ALICE \TPC\ is a large cylindrical drift detector
whose active volume extends radially from 85 to 247 cm, and from -250
to +250 cm along the beam direction. The active volume of nearly 90
m$^3$ is filled with a gas mixture of Ne (85.7\%), CO$_2$ (9.5\%) and
N$_2$ (4.8\%) until the end of 2010, and Ne (90\%) and CO$_2$ (10\%)
since the beginning of 2011.  A central electrode maintained at
-100~kV divides the \TPC\ into two sections. The end-caps are equipped
with multiwire proportional chambers with cathode pad readout. For a
particle traversing the \TPC\, up to 159 position signals (clusters)
are recorded.  The cluster data are used to reconstruct the
charged particle trajectory as well as to calculate the particle’s
specific energy loss used to identify the species of the particle
which has produced the track.

The two Zero Degree Calorimeters (\ZDC) in the ALICE experiment
measure the energy of spectator (non-interacting) nucleons: ZP
measures protons and ZN measures neutrons. They are situated about
114~m from the interaction point on each side of the experiment
\cite{aliceapp}.  Each \ZDC\ consists of two quartz fiber sampling
calorimeters: the neutron calorimeter positioned between the two beam
pipes downstream of the first machine dipole that separates the two
charged particle beams, and the proton calorimeter positioned
externally to the outgoing beam pipe.  The energy resolution at beam
energy is estimated to be 20\% for the neutron (20.0\% for ZNC, 21.2\%
for ZNA) and 24\% for the proton calorimeters, respectively.

\begin{figure}[thb!f]
 \centering
 \includegraphics[width=0.6\textwidth]{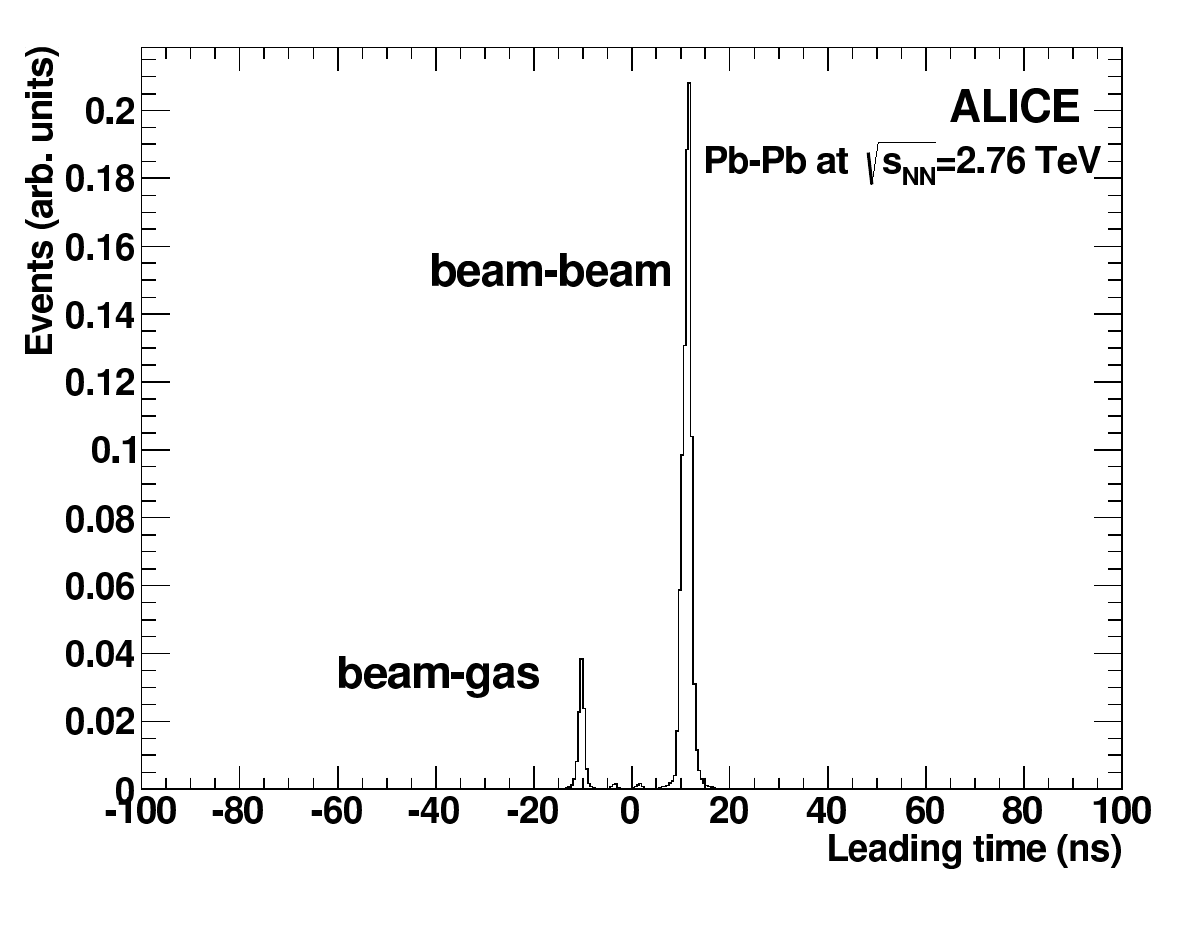}
 \caption{Time distribution of signals in the VZERO detector on the A
  side. The peaks corresponding to beam-beam, beam-gas and satellite
  collision events are clearly visible.
  \label{fig:v0_beam_gas}}
\end{figure}

\subsection{Data set and online event selection}
\label{subsec:event-selection}

During the first LHC \PbPb\ run in 2010, beams of four bunches with
about $10^7$ Pb ions per bunch collided at $\snn=2.76$~TeV, with an
estimated luminosity of $5 \times 10^{23} \rm cm^{-2} \rm s^{-1}$.
ALICE collected about 90 million nuclear collision events using
different interaction triggers with increasingly tighter conditions.
These triggers used \VZERO\ and \SPD\ detector signals in coincidence
with a bunch crossing corresponding to a beam-beam collision:

\begin{itemize}
\item \emph{V0AND}: signals in \VZEROA\ and \VZEROC;
\item \emph{3-out-of-3}: signals in \VZEROA\ and \VZEROC\ and at least 2 chips
  hit in the outer layer of the \SPD;
\item \emph{2-out-of-3}: two of the three conditions listed above.
\end{itemize}

The threshold in the \VZERO\ detector for each of the \VZERO\ tiles
corresponded approximately to the energy deposition of one minimum
ionizing particle.

Control events were also collected with the same trigger logic, in
coincidence with only one beam crossing the ALICE interaction point
(from either the A or the C side) or with no beam at all (``empty'').
The luminous region had an RMS width of 5.9~cm in the longitudinal
direction and 50~$\mu$m in the transverse direction.  For the
estimated luminosity, using the least selective of the interaction
triggers, the observed rate was about 50 Hz. This was mainly due to
electromagnetically induced processes \cite{Baur:2001jj}.  These
processes have large cross sections at LHC energies but generate
low multiplicities and therefore do not contribute to the typical
particle multiplicities of interest for the present paper.  
The trigger rate without beam was negligible and the rate
in coincidence with bunches of only one beam was about 1~Hz.  The
probability for collision pile-up per triggered event was less than
$10^{-4}$.

\begin{figure}[htb!f]
 \centering
 \includegraphics[width=0.6\textwidth]{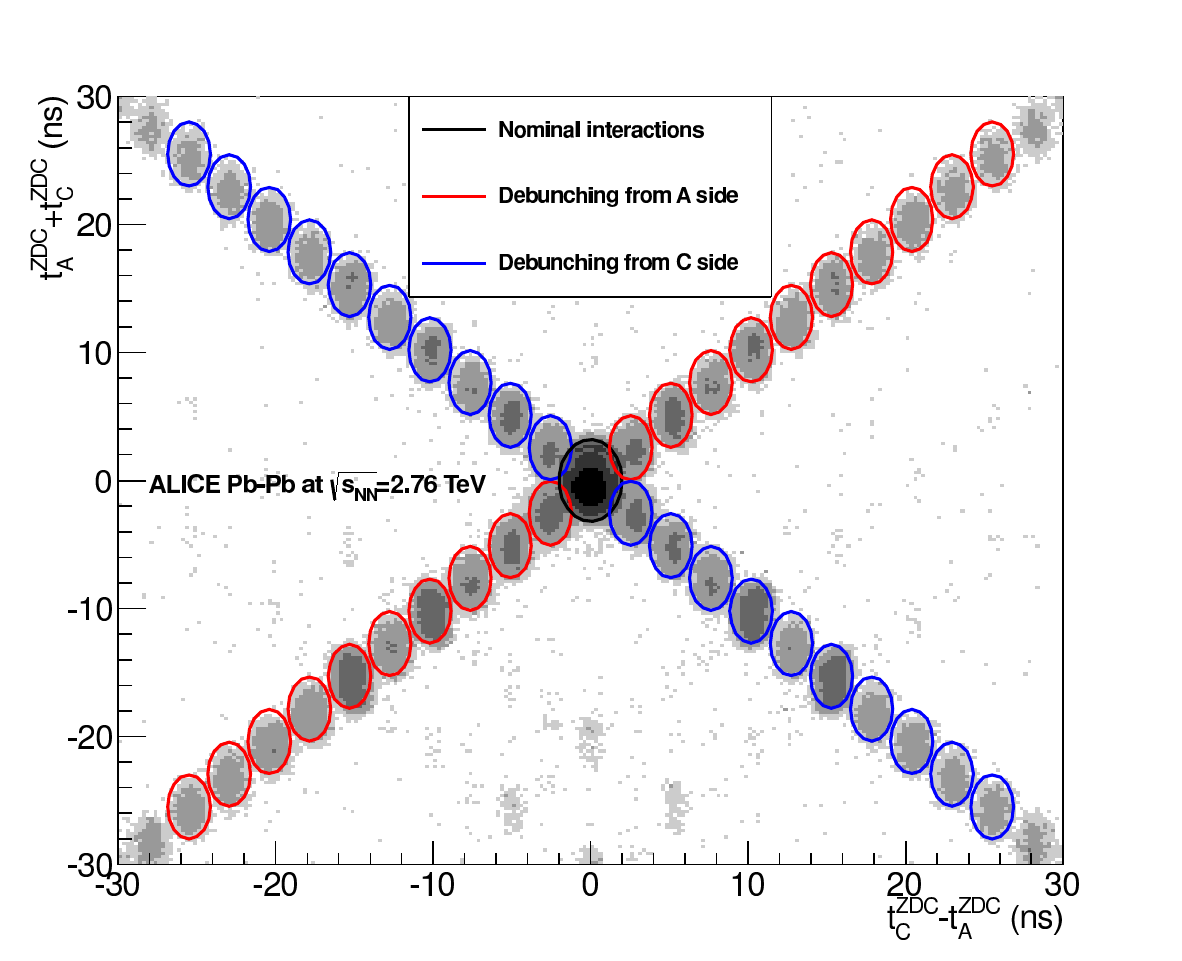}
 \caption{(Color online) Correlation between the sum and the difference of times
  recorded by the neutron \ZDC\ on either side of the interaction
  region. The large cluster in the middle corresponds to collisions
  between ions in the nominal RF buckets of each beam, while the
  small clusters along the diagonals (spaced by 2.5~ns in the time
  difference) correspond to collisions in which one of the ions is 
  displaced by one or more RF buckets.
  \label{fig:zdc_debunch}}
\end{figure}

\subsection{Offline event selection}
\label{subsec:offselection}

The offline event selection is applied with the purpose of selecting
hadronic interactions with the highest possible efficiency, while
rejecting the machine-induced and physical backgrounds.  The offline
event selection replays the on-line trigger condition, using the same
quantities calculated offline, so that events triggered by noise in
the \SPD\ are discarded, and the weighted time average over all 
channels is used for the \VZERO, leading to a better time resolution.  
In addition, the
offline event selection rejects the machine-induced background and
parasitic collisions. This contamination amounts to about 25\% of all
collected events. To keep the conditions of all detectors as uniform
as possible (in particular those around mid-rapidity, such as the
\SPD), the centrality analysis was restricted to a region around the
vertex, $\left|z_{vtx}\right|\lesssim 10~\mathrm{cm}$.

\subsubsection{Machine-Induced Background}
\label{subsubsec:machine-bkg}

One source of machine-induced background is due to beam-gas events,
caused by one of the beams interacting with the residual gas in the
beam-pipe; another source of background are events where ions in the
beam halo interact with mechanical structures in the machine. These
interactions mostly occur outside of the interaction region and thus
produce a signal that is "too early" in the same-side \VZERO, compared
to a collision that occurs in the nominal interaction region between
the \VZERO\ detectors. Therefore these events can be rejected using
the timing information of the \VZERO. This is illustrated in
\Fig{fig:v0_beam_gas} which shows the arrival time of particles at the
\VZERO\ A detector relative to the nominal beam crossing
time. Beam-halo or beam-gas interactions are visible as secondary
peaks in the time distribution because particles produced in
background interactions arrive at earlier times in the detector
relative to particles produced in beam-beam collisions at the
nominal vertex, which are the majority of the signals.  Other small
peaks between these main ones arise from satellite collisions.

Another source of machine-induced background is due to parasitic
collisions from debunched ions. The radio-frequency (RF) structure of
the LHC of 400 MHz is such that there are 10 equidistant RF buckets
within the 25~ns time interval between two possible nominal bunch positions.
Therefore the buckets are spaced by 2.5~ns. Only one of them should be
populated by ions~\cite{Evans:2008zzb}. However, ions can ``jump'' 
into one of the neighboring buckets. Therefore
collisions occur either between ions in the nominal RF buckets but
also between one or two ions displaced by one or more RF buckets. This
causes a displacement in the Z-vertex position of
$2.5~\mathrm{ns}/2\cdot c = 37.5~\mathrm{cm}$, well outside the
fiducial region $\left|z_{vtx}\right|\lesssim 10~\mathrm{cm}$. Those
events are thus to be considered as ``background'' and are rejected
using the correlation between the sum and the difference of times
measured in each of the neutron ZDCs, as shown in \Fig{fig:zdc_debunch}. 
Such satellite collisions can also be rejected using the vertex cut.

After the event selection, the remaining machine-induced background,
estimated from the control triggers (i.e. triggers that fire for
coincidences between empty and filled or empty and empty bunches), is
negligible.

\begin{figure}[htb!f]
\centering
\includegraphics[width=0.6\textwidth]{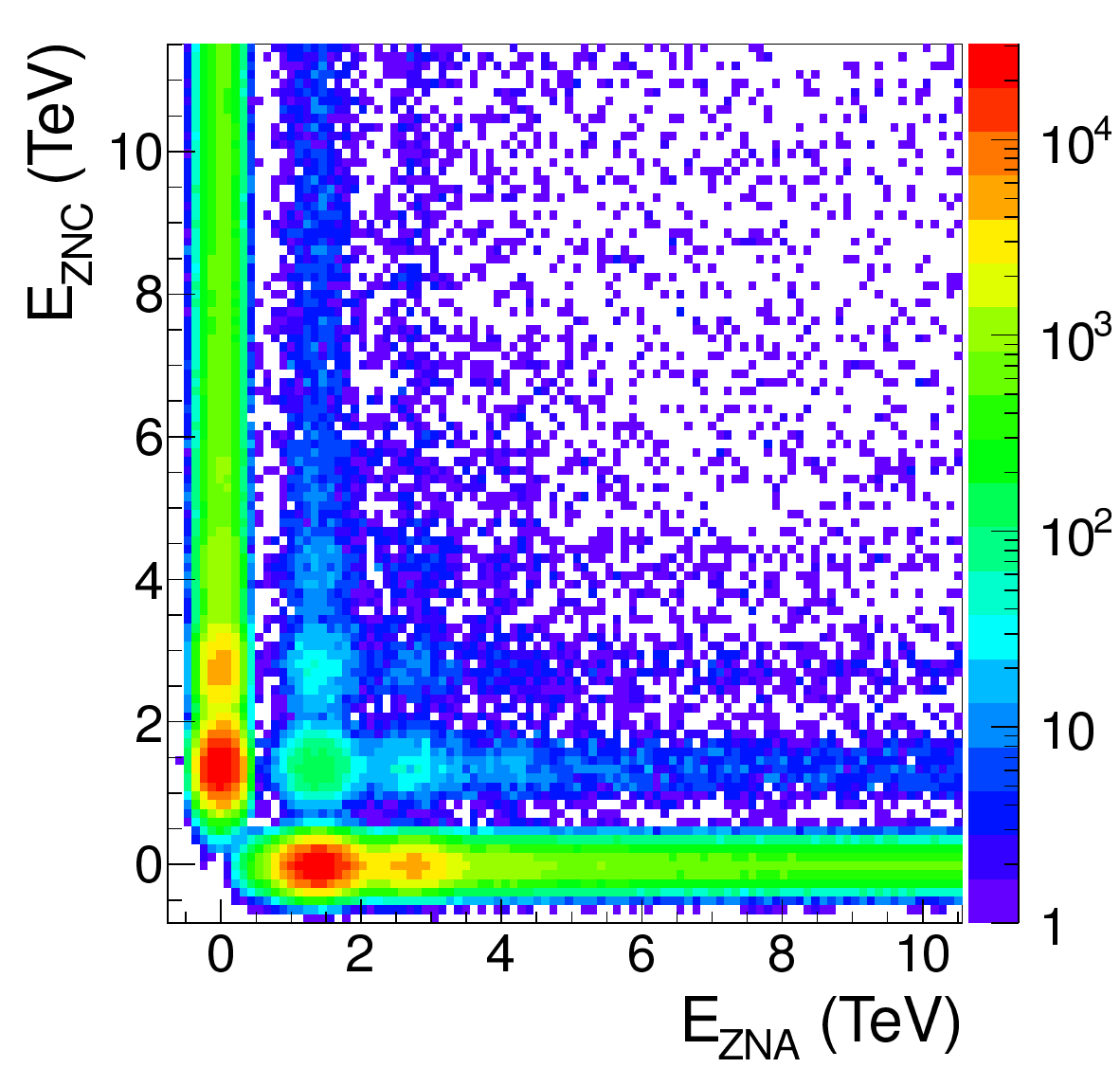}
\caption{(Color online) Correlation between signals in the two neutron zero-degree
 calorimeters, ZNA and ZNC. The figure is taken from \cite{Oppedisano:2011}.
 Single electromagnetic dissociation events produce signal in only one of the 
 calorimeters.  Mutual dissociation and hadronic interactions populate interior 
 of the plot and can be distinguished from each other by the signal in ZEM.
 \label{fig:zncvszna}}
\end{figure}

\subsubsection{Electromagnetic interaction background} 
\label{subsubsec:EM} 

At the LHC energy, the cross sections for electromagnetic (EM)
processes, generated by the EM fields of relativistic heavy ions, are
enormous ($\mathcal{O}(\mathrm{kbarn})$)
\cite{Bruce2009,Djuvsland:2010qs,HEN03-QED,Oppedisano:2011}.  This is
the main physical background, and needs to be rejected in heavy-ion
collisions to isolate hadronic interactions.  QED processes consist
of: photo-production and photo-nuclear interactions. Photo-production
results in the creation of an $e^+e^-$ pair. Photo-nuclear
interactions, where one photon from the EM field of one of the nuclei
interacts with the other nucleus, possibly fluctuating to a vector
meson, yield a low multiplicity of soft particles in the ALICE central
barrel.  In the case of single photo-production the particle
multiplicity is asymmetric within the event.  Along the beam direction
the electromagnetic dissociation (EMD) cross sections are large resulting 
in a non-negligible probability for one neutron emission from either nucleus.

The EMD cross sections have been measured in a special run triggering on a signal 
in one of the neutron ZDCs, ZNA or ZNC, with a threshold  placed well below the 
single neutron signal to detect the neutrons from Giant Dipole Resonance (GDR) 
decay emitted very close to beam rapidity\cite{Oppedisano:2011}.
The recorded event sample is dominated by electromagnetic dissociation of one or 
both nuclei measured to be $\sigma^{\rm single\ EMD} = 187.4 \pm 0.2 (stat.) ^{+13.2}_{-11.2} (syst.)$~b 
compared to the mutual EMD cross section of $\sigma^{\rm mutual\ EMD} = 5.7 \pm 0.1 (stat.) \pm 0.4 (syst.)$~b. 
The single EMD events can be clearly identified when correlating the response of ZNA and ZNCs~(Fig.~\ref{fig:zncvszna}).  
The additional requirement of a signal in an electromagnetic calorimeter close to beam rapidity (ZEM) allows one to distinguish
between mutual EMD and hadronic interaction events.

In order to reduce the contribution due to single neutron emission, we require a \ZDC\ signal 
three standard deviations above the single neutron peak. This selection rejects about 3\% of 
all events within 10 cm from the nominal interaction point after removal of beam-gas and
parasitic collisions, and only removes events for peripheral collisions (in the 90--100\% region).  
The coincidence of the \ZDC\ signals rejects all the single neutron emission events. 
The simultanous emissions from both nuclei still are accepted, which, however, are only relevant 
for very peripheral collisions. 

For systematic studies, another selection based on the information from the \TPC\ is
used, where at least one track reconstructed in the \TPC\ is requested in order to keep 
the event. This selection removes few peripheral hadronic interactions, and strongly
suppresses the EM background.

\section{Determination of the hadronic cross section}
\label{sec:mult}

In order to classify the collisions in percentiles of the hadronic
cross section using the charged particle multiplicity, it is necessary
to know the particle multiplicity at which the purity of the event
sample and the efficiency of the event selection becomes 100\%. 
We define the \emph{Anchor Point} (AP) as the amplitude of the \VZERO\
detector equivalent to 90\% of the hadronic cross section, which
determines the absolute scale of the centrality.
The determination of the AP requires the knowledge of the
trigger efficiency and the remaining background contamination in
nuclear collision events. Two methods have been used to study this.
The difference in the results obtained with the two methods is used to
estimate the systematic uncertainty by:

\begin{itemize}
\item
{\bf Simulating the multiplicity distribution} (see
Sec.~\ref{subsec:correctmult}).  In the first approach, we use a full
simulation of hadronic and EM processes, including a detailed
description of the detector response, to study the efficiency of the
event selection (Sec.~\ref{subsubsec:trigeff}) and to estimate the
background contamination (Sec.~\ref{subsubsec:trigpurity}). The real
multiplicity distribution, corrected for efficiency and purity, allows
direct access to the AP.
\item
{\bf Fitting the multiplicity distribution} (see
Sec.~\ref{subsec:NBD-glauber}).  In the second method, we use the
Glauber Monte Carlo, combined with a simple model for particle
production, to simulate a multiplicity distribution which is then
compared to the experimental one. The simulated distribution describes
the experimental one down to the most peripheral events where they start
to deviate due to background contamination and limited trigger
efficiency. The location of the divergence between the data and simulation
can be used to define the AP.
\end{itemize}

The centrality determination is performed for different trigger and detector settings.  
The different triggers change the fraction of accepted events from EM processes, 
which determines the shape of the multiplicity distribution for very peripheral 
collisions below the AP.
The position of the AP is very stable for the entire 2010 (and 2011) run period and 
does not change within the quoted systematic uncertainty discussed below.
Small variations in detector conditions induce small changes in the position 
of the edge of the multiplicity distribution for most central events.  
Nevertheless, the centrality determination, adjusted to account for small changes 
in the detector configuration, provides a stable centrality selection for the entire 
data-taking period (the mean fraction of events in the 0-1\% bin is 0.01 with a RMS of 0.001, 
and in the 40-50\% bin the mean fraction is 0.101 with an RMS of 0.002).

\begin{figure}[t]
 \centering
 \includegraphics[width=0.75\textwidth]{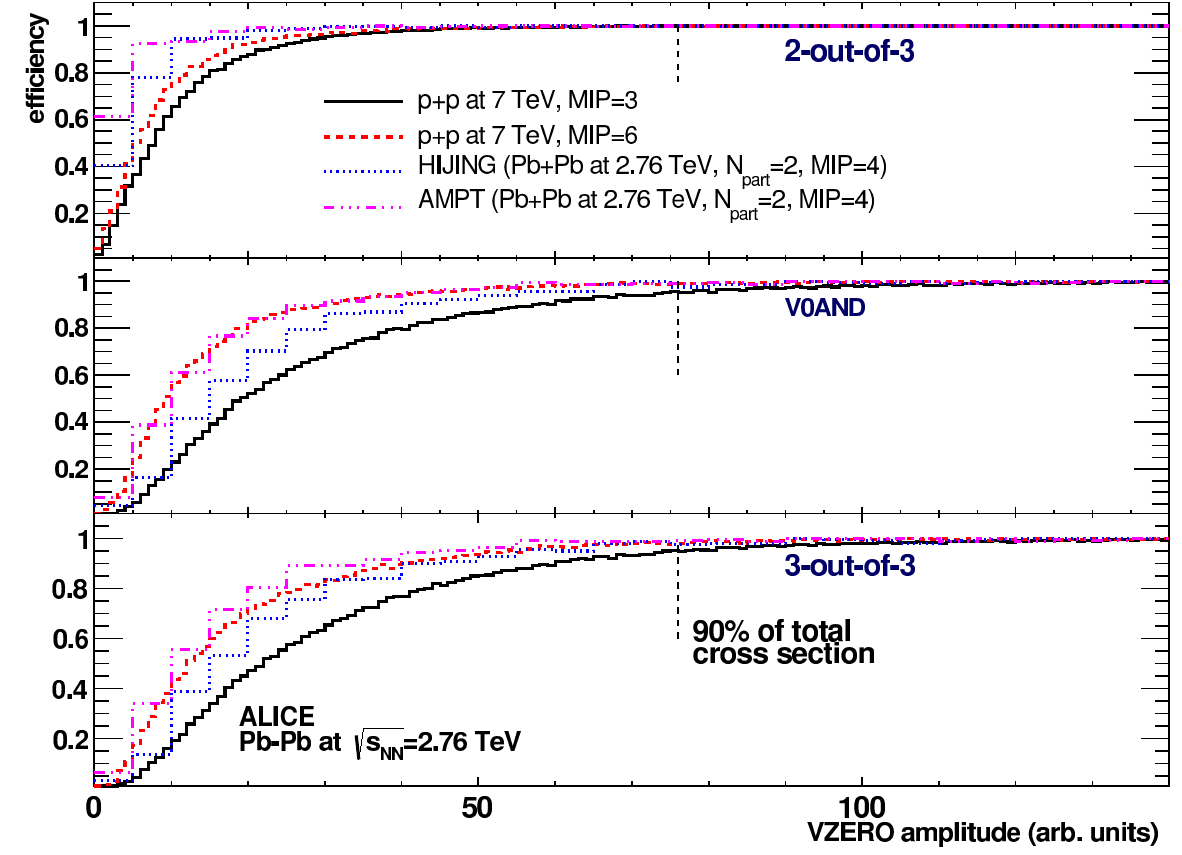}
 \caption{(Color online) Efficiency of the three online triggers
  ($2\mathrm{-out-of-}3$, \emph{V0AND}, $3\mathrm{-out-of-}3$) used
  for \PbPb\ collisions as a function of the \VZERO\ amplitude
  calculated with HIJING and AMPT, and measured in dedicated \pp\ runs.
  The efficiency in the simulation has been calculated for events with
  $\Npart=2$.
  \label{fig:trieff}}
\end{figure}

\subsection{Method 1: Correcting the multiplicity distribution}
\label{subsec:correctmult}
With this method the AP is determined by evaluating the efficiency of the event selection 
and by estimating the purity of the obtained event sample.

\subsubsection{Efficiency of the event selection}
\label{subsubsec:trigeff}

The efficiency for the different event selections is studied with
simulations of hadronic reactions and with dedicated \pp\ runs.  For
simulations we use HIJING \cite{hijing} or AMPT \cite{ampt} with a
full GEANT \cite{geant3ref2} description of the ALICE detector and a
trigger emulator.  
In the simulations the efficiency is defined as the ratio of events
selected by a given condition to all generated events. In the two dedicated
\pp\ runs, which were taken at the end of the 2010 run at \sonly\
=~7~TeV, the detector conditions were similar to those in the \PbPb\ run. 
The \VZERO\ gain was adjusted such that the response to minimum ionizing 
particles (MIP) corresponded to 3 ADC channels for one run, or 6 for the other 
run.  For the \PbPb\ run, it was set to 4 channels, i.e.\ between the two tested 
conditions.
In the special \pp\ runs, we used a minimum interaction trigger,
which requires a logical OR between a hit in the \SPD\, and in either of
the two \VZERO\ detectors (CINT1 trigger condition). 
The same event selection criteria as used in the \PbPb\ run as the trigger 
have been applied. The relative event selection efficiency is defined as 
the ratio of events selected by a given condition to all the 
events recorded with the \pp\ minimum-bias interaction trigger (CINT1). 
Since the \pp\ minimum-bias interaction trigger (CINT1) has an efficiency that 
is effectively 100\% for non-diffractive events~\cite{Poghosyan:2011}, 
the relative efficiency measured in the \pp\ runs, shown in Fig.~\ref{fig:trieff}, 
can be qualitatively compared to that obtained in \PbPb\ simulations with \Npart\ = 2.
Except for very low amplitudes, results from HIJING and AMPT are in very good agreement.  
AMPT predicts a slightly higher efficiency (about 0.5\%), as a consequence of the broader 
rapidity distribution. The comparison with the \pp\ runs shows a reasonable agreement for the
``MIP = 6'' case, while the ``MIP = 3'' is clearly lower.

For the \PbPb\ run, the efficiency of the event selection is
calculated using the average of results obtained with HIJING and AMPT.
The efficiency of the interaction triggers is 99.4\%, 97.1\%, 96.9\%
respectively for \emph{2-out-of-3}, \emph{V0AND},
\emph{3-out-of-3} using HIJING and 99.7\%, 98.6\%, 98.4\% using AMPT.  
The line in Fig.~\ref{fig:trieff}, corresponding to the 90\% of the
hadronic cross section, shows that the trigger is always fully
efficient for the 90\% most central collisions, except for the ``MIP = 3''
\pp\ case, where the efficiency is 95\%.

\begin{figure}[thb!f]
 \centering
 \includegraphics[width=0.8\textwidth]{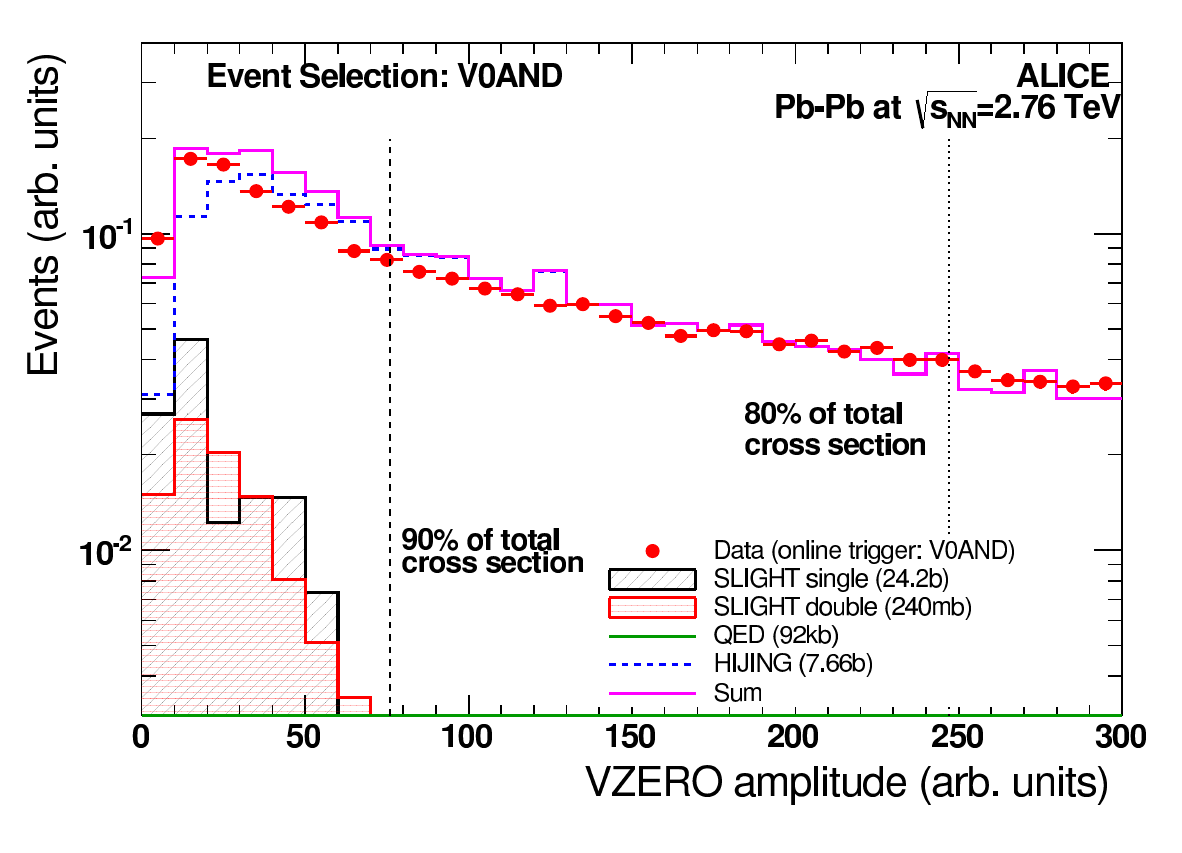}
 \caption{(Color online) \VZERO\ amplitude distribution in data (red points) and
  simulations with the V0AND interaction trigger. The data are
  compared to the sum of HIJING + QED + STARLIGHT simulations
  (histogram) with the same event selection.
 \label{fig:v0m_datasim}}
\end{figure}

\subsubsection{Remaining contamination}
\label{subsubsec:trigpurity}

The purity of the data sample passing a given event selection is
estimated using HIJING simulations \cite{hijing} for hadronic
processes and QED \cite{HEN03-QED} and STARLIGHT
\cite{Djuvsland:2010qs} for the simulations of the EM background.  
For the electromagnetic dissociation we assume that the selection based on
the signal 3$\sigma$ above the single-neutron peak in the ZDCs (see~\ref{subsubsec:EM}) 
is  fully efficient. 

In \Fig{fig:v0m_datasim}, data taken with the V0AND interaction trigger are
compared to the sum of HIJING and background (QED + STARLIGHT)
simulations with the same event selection. The simulations are scaled
to the known cross sections:
\begin{itemize}
\item HIJING (hadronic): $\sigma_{H}$ = 7.66 b \cite{hijing};
\item QED (EM): $\sigma_{Q}$ = 92 kb \cite{HEN03-QED}; 
\item STARLIGHT (single neutron dissociation): $\sigma_{SNS}$ = 24.2 b \cite{Djuvsland:2010qs}; 
\item STARLIGHT (double neutron dissociation): $\sigma_{SND}$ = 240 mb \cite{Djuvsland:2010qs}. 
\end{itemize}
The sum of the simulations is normalized to the data in the region
150 $<$ \VZERO\ amplitude $<$ 500, where there is no background
contamination. The contribution from QED is completely removed by the
V0AND trigger. The dashed lines, indicating respectively 80\% and 90\%
of the hadronic cross section, show that there is no significant
background contamination for collisions more central than 90\%.  The
region 90--100\% is reasonably understood as the agreement between data
and simulation is quite good. The remaining discrepancy between the data 
and the sum of all contributions is included in the systematic uncertainty.

To assign a systematic uncertainty, the comparison is made for the
three online interaction triggers and other event selections requiring
(i) V0AND + \TPC: one track fully reconstructed in the \TPC\ on top of
the V0AND trigger; (ii) V0AND + \ZDC: 3$\sigma$ cut above single
neutron peak in \ZDC\ on top of the V0AND trigger. For all these
variations a cross section is calculated and the difference is
included in the systematic uncertainty.

\begin{figure}[thb!f]
 \centering
 \includegraphics[width=0.65\textwidth]{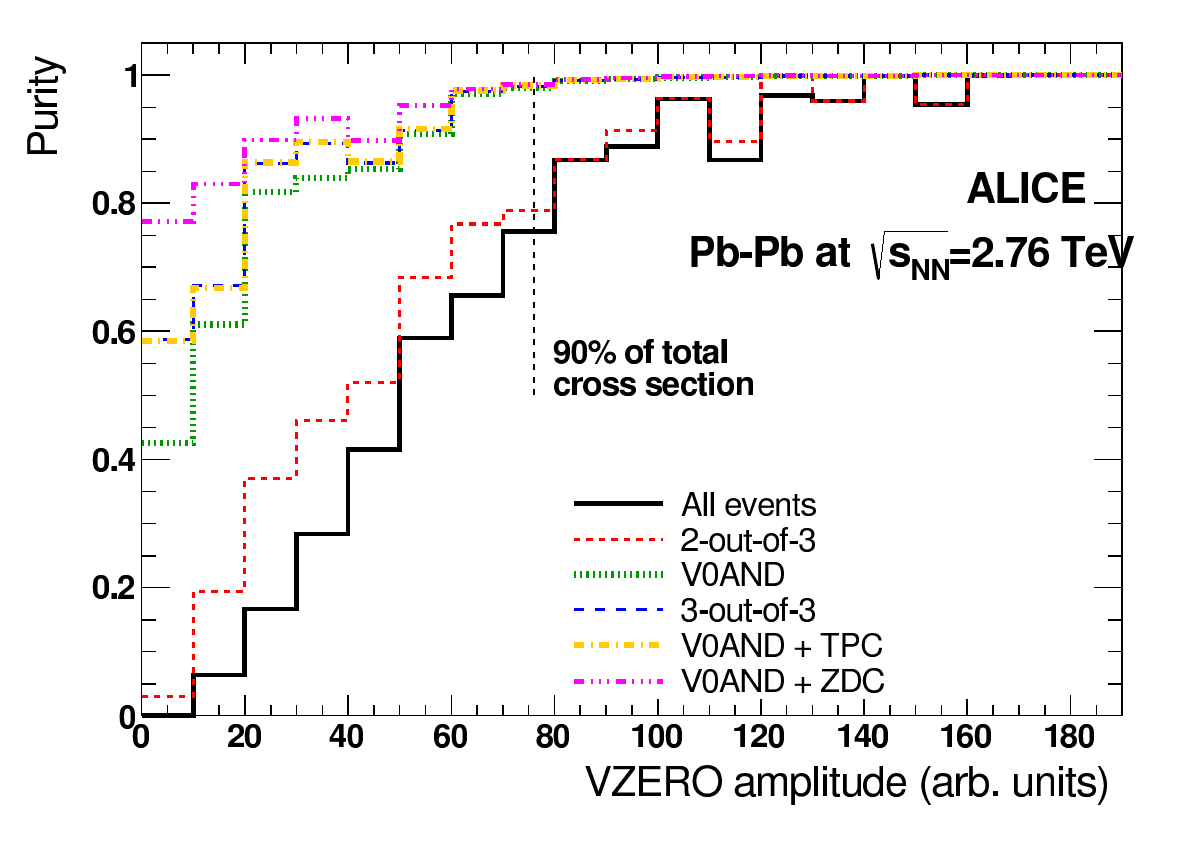}
 \caption{(Color online) Purity of the 3 online interaction triggers
  (\emph{2-out-of-3}, \emph{V0AND}, \emph{3-out-of-3}) and other event selections
  used for \PbPb\ collisions as a function of the \VZERO\ amplitude
  calculated with HIJING, STARLIGHT and QED simulations. The dashed
  line indicates 90\% of the hadronic cross section.
  \label{fig:purity}}
\end{figure}

\Figure{fig:purity} shows the purity of the various \PbPb\ event
samples after those selections.  The purity, plotted as a function of
the \VZERO\ amplitude (V), is defined as the fraction of hadronic
collisions over all the events selected with a given condition:
\begin{equation} 
\mathrm{purity} = \frac{{\frac{\rm{dN_{x}}}{\rm{d}V}|_H} \frac{\sigma_H}{N_H} } 
{{\frac{\rm{dN_{x}}}{\rm{d}V}|_{H}}  \frac{\sigma_{H}}  {N_{H}} +
 {\frac{\rm{dN_{x}}}{\rm{d}V}|_{SNS}} \frac{\sigma_{SNS}}{N_{SNS}} +
 {\frac{\rm{dN_{x}}}{\rm{d}V}|_{SND}} \frac{\sigma_{SND}}{N_{SND}} +
 {\frac{\rm{dN_{x}}}{\rm{d}V}|_{Q}}   \frac{\sigma_{Q}} {N_{Q}} }.
\end{equation} 

where $\sigma_x$ and $N_x$ are the cross sections and number of events
for a given process, $x$, where $x=H$, $SNS$, $SND$, and $Q$, for
HIJING, STARLIGHT single, STARLIGHT double, and QED, respectively.

The purity of the event sample can be verified using the correlation of
the energy deposition in the two sides of the ZN calorimeter, similar
to the one shown in Fig.\ref{fig:zncvszna}. Single neutron peaks are visible
in the 80--90\% centrality class, which may indicate some remaining
contamination from EMD events. However their origin can be also
attributed to asymmetric \PbPb\ events, as well as a pile up of an EMD
and a hadronic collision. Since this contamination can not be easily
removed, analyses that use peripheral classes like 80--90\% assign an
additional 6\% systematic uncertainty on the event selection to take
into account the possible contamination from EMD.

\begin{figure}[ht]
 \centering
 \includegraphics[width=0.75\textwidth]{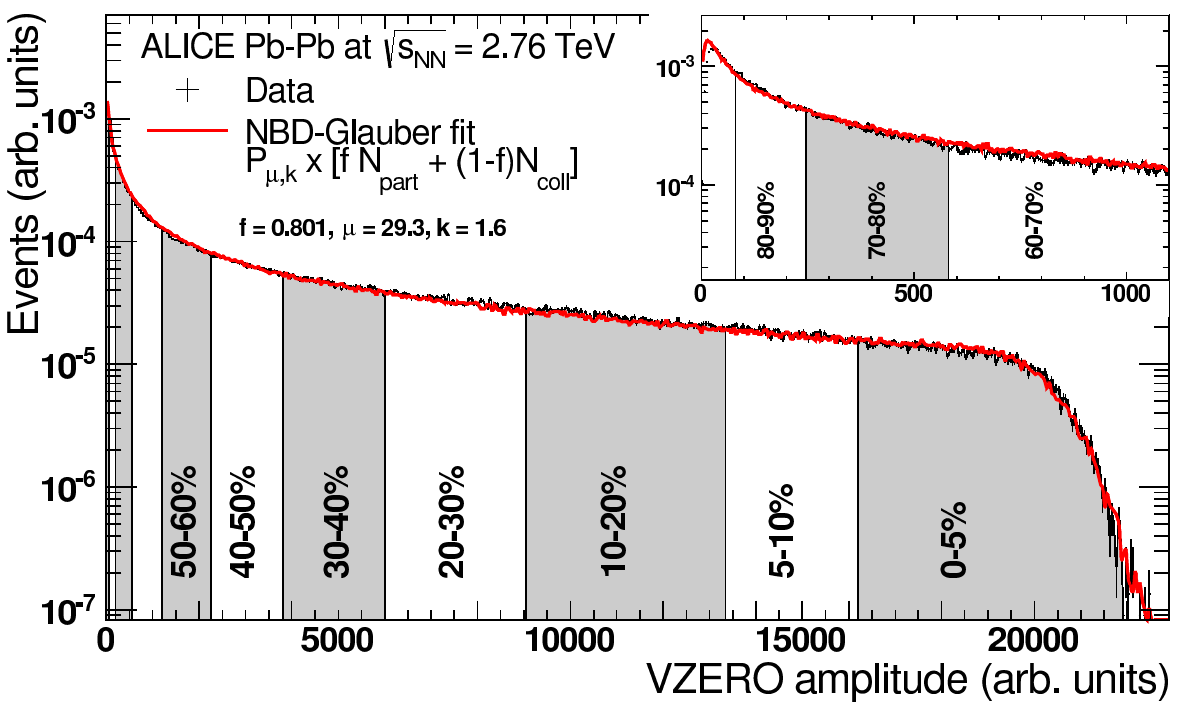}
 \caption{(Color online) Distribution of the sum of amplitudes in the \VZERO\
  scintillators. The distribution is fitted with the NBD-Glauber fit
  (explained in the text) shown as a line. The centrality classes
  used in the analysis are indicated in the figure. The inset shows
  a zoom of the most peripheral region.
  \label{fig:glau}}
\end{figure}

\subsection{Method 2: Fitting the multiplicity distribution}
\label{subsec:NBD-glauber}

Another independent way to define the AP uses a
phenomenological approach based on the Glauber Monte Carlo to fit the
experimental multiplicity distribution.  The Glauber Monte Carlo uses
the assumptions mentioned above plus a convolution of a model for
particle production, based on a negative binomial distribution (NBD).
This latter assumption is motivated by the fact that in minimum bias
\pp\ and $\ppbar$ collisions at high energy, the charged particle
multiplicity $d\sigma/dN_\mathrm{ch}$ has been measured over a wide
range of rapidity and is well described by a NBD~\cite{alicepp1,alicepp2}. 
This approach allows one to simulate an experimental multiplicity distribution 
(e.g.\ \VZERO\ amplitude), which can be compared with the one from data.

\Figure{fig:glau} shows the distribution of \VZERO\ amplitudes for all
events triggered with the \emph{3-out-of-3} trigger~(see
\ref{subsec:event-selection}) after removing the beam background~(see
\ref{subsubsec:machine-bkg}), part of the EM background with the \ZDC\
cut (see \ref{subsubsec:EM} ) and a Z-vertex cut $|z_{vtx}|<10$~cm.
The multiplicity distribution has the classical shape of a peak corresponding to most
peripheral collisions (contaminated by EM background and by missing
events due to the trigger inefficiency), a plateau of the intermediate
region and an edge for the central collisions, which is sensitive to
the intrinsic fluctuations of \Npart\ and $\dNdeta$ and to detector
acceptance and resolution.  

The Glauber Monte Carlo defines, for an event with a given impact
parameter $b$, the corresponding \Npart\ and \Ncoll.  The particle
multiplicity per nucleon-nucleon collision is parametrized by a NBD.
To apply this model to any collision with a given \Npart\ and \Ncoll\
value we introduce the concept of ``ancestors'', i.e.\ independently 
emitting sources of particles.  
We assume that the number of ancestors \Nanc\ can be parameterized by 
$\Nanc = f\cdot\Npart+(1-f)\cdot\Ncoll$. 
This is inspired by two-component models \cite{Kharzeev:2004if,Deng:2010mv}, 
which decompose nucleus--nucleus collisions into soft and hard interactions, 
where the soft interactions produce particles with an average multiplicity 
proportional to \Npart, and the probability for hard interactions to occur is 
proportional to \Ncoll. We discuss the independence of the fit results of this 
assumption below (Sec.~\ref{subsubsec:ancestors}).

To generate the number of particles produced per interaction, we use 
the negative binomial distribution
\begin{equation}
  P_{\mu,k}(n) =  \frac{\Gamma(n+k)}{\Gamma(n+1)\Gamma(k)} \cdot
  \frac{(\mu/k)^n}{(\mu/k+1)^{n+k}}\,,
\end{equation}
which gives the probability of measuring $n$ hits per ancestor, where
$\mu$ is the mean multiplicity per ancestor and $k$ controls the
width.  For every Glauber Monte Carlo event, the NBD is sampled \Nanc\
times to obtain the averaged simulated \VZERO\ amplitude for this
event, which is proportional to the number of particles hitting the
hodoscopes.  The \VZERO\ amplitude distribution is simulated for an
ensemble of events and for various values of the NBD parameters $\mu$,
$k$, and the \Nanc\ parameter $f$. A minimization procedure is applied
to find the parameters which result in the smallest $\chi^2$, also
shown in Fig.~\ref{fig:glau}.  The fit is performed for \VZERO\
amplitudes large enough so that the purity of the event sample and the
efficiency of the event selection is 100\%.  That leaves a very broad
range in the amplitude values that can be fitted to extract parameters
$f$, $\mu$ and $k$ directly from the data. The amplitude, above which
we have 90\% of the hadronic cross section, defines the AP.  The
quality of the fit is good, as the $\chi^2$/NDF is approximately unity
for all fits. We note that the high multiplicity tail, which is quite
sensitive to fluctuations and the detector resolution not implemented
in the model, is not perfectly well described. Even replacing the
black-disk nucleon-nucleon overlap function with a Gaussian does not
improve the fit, as the difference in the \Npart\ distribution is
washed out in the \Nch\ distribution.  However, it is important to
remark that the fit is used solely to determine the AP, which is quite
insensitive to the detailed shape of the high multiplicity tail.

An equivalent procedure was applied to fit, with the NBD-Glauber
method, the distribution of the hits collected in the outer layer of
the \SPD, and the tracks reconstructed in the \TPC. All these analyses
give consistent results, which are summarized in
Sec.~\ref{subsec:sysAP}.

\begin{figure}[ht]
 \centering
 \includegraphics[width=0.75\textwidth]{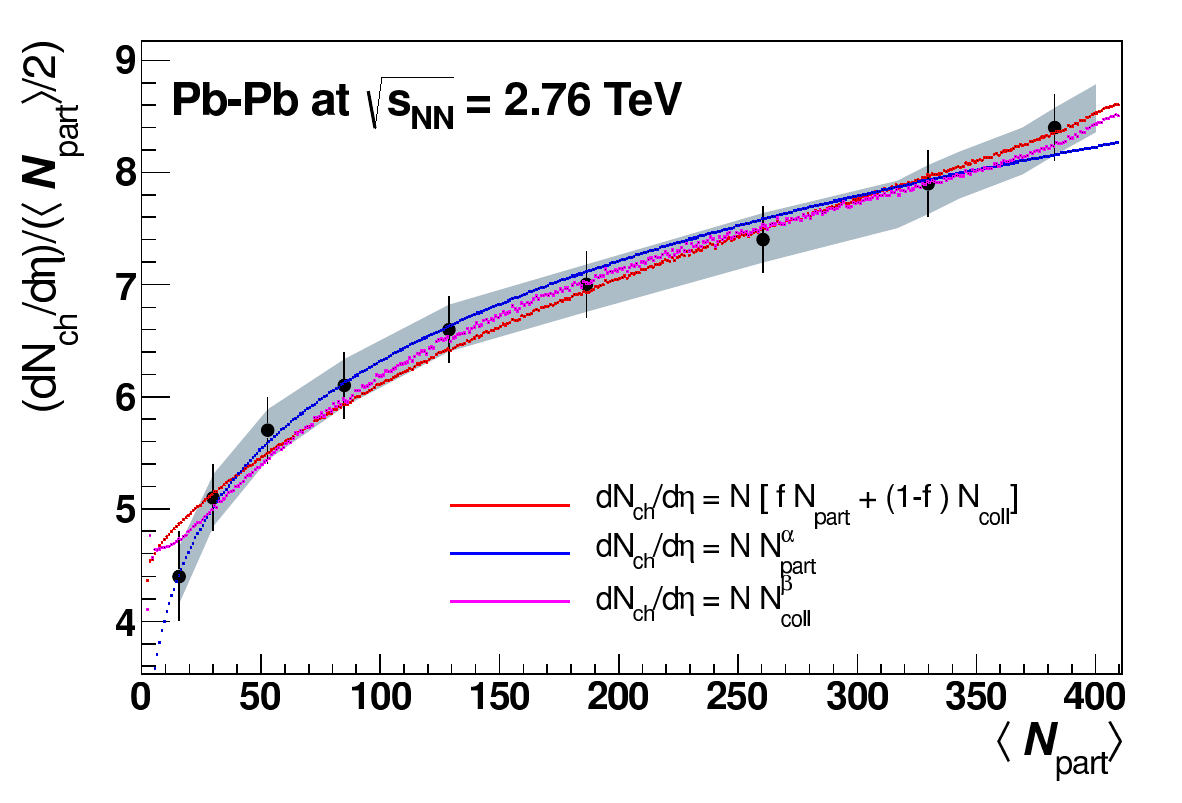}
  \caption{(Color online) Centrality dependence of $\dNdeta$ per participant
   pair as a function of $\Npart$, measured in the \PbPb\ data at
   $\snn = 2.76$~TeV fitted with various parametrizations of
   $\Npart$ and $\Ncoll$, calculated with the Glauber model. The
   fit parameters are given in the figure. Data are from~\cite{alicecent}.
  \label{fig:ancestors}}
\end{figure}

\subsubsection{Ancestor dependence} 
\label{subsubsec:ancestors} 

The number of emitting sources \Nanc\ is determined by a function
inspired by the two-component models, i.e.\ $\Nanc =
f\cdot\Npart+(1-f)\cdot\Ncoll$.  However, other assumptions can be made
leading to a different parametrization, which are briefly discussed in
the following.
The ancestor dependence on \Npart\ and \Ncoll\ derives from a parametrization 
of the dependence of the charged particle multiplicity on \Npart\ and \Ncoll. 
Systematic studies of this dependence performed at the SPS \cite{Abreu:2002fw,
  Aggarwal:2001, Antinori:2001qn}, at RHIC \cite{Adler2005}, and recently at the LHC
\cite{alicecent, ATLASmult, CMSmult}, have been used in an attempt to
constrain different models of particle production.

\begin{table*}[thb!f]
 \caption{Parameters of the fit to the charged particle multiplicity for the three different 
  parametrizations discussed in the text, with error and $\chi^2$/NDF.
  \label{tab:FitPar}}
\begin{tabular}{c|ccccc}
Model & Normalization & Error & Fit Par. & Error & $\chi^2$/NDF \\
\hline
$f \cdot \mathrm{N_{\mathrm{part}}} + (1-f) \cdot \mathrm{N_{\mathrm{coll}}}$ & 2.441 & 0.281 & $f     $~=~0.788 & 0.021 & 0.347 \\
$\mathrm{N_{\mathrm{part}}}^{\alpha}$                                        & 1.317 & 0.116 & $\alpha$~=~1.190 & 0.017 & 0.182 \\
$\mathrm{N_{\mathrm{coll}}}^{\beta}$                                         & 4.102 & 0.297 & $\beta $~=~0.803 & 0.012 & 0.225 \\
\end{tabular}
\end{table*}

The charged particle multiplicity is expected to scale with \Npart\ in
scenarios dominated by soft processes. In this case, all the
participant nucleons can be assumed to contribute with the same amount
of energy to particle production, and the scaling with \Npart\ is
approximately linear. By contrast, a scaling with \Ncoll\ is expected
for nuclear collisions in an energy regime where hard processes
dominate over soft particle production. In this case, nuclear
collisions can be considered as a superposition of binary
nucleon-nucleon collisions.  Two-component models are used to quantify
the relative importance of soft and hard processes in the particle
production mechanism at different energies.

To determine the scaling behavior of the particle production, the
charged particle multiplicity $\dNdeta$ as a function of the number of
participants \Npart\ was fitted with a power-law function of \Npart\,
i.e. $\dNdeta \propto \Npart^{\alpha}$.  While at SPS energy the
scaling with \Npart\ is approximately linear, i.e.\ $\alpha \sim 1$
\cite{Abreu:2002fw, Aggarwal:2001, Antinori:2001qn}, results from the experiments at
RHIC show evidence of a large contribution of hard processes to
particle production, resulting in $\alpha > 1$.

The charged particle multiplicity per participant pair \dNdetapt\
measured by ALICE \cite{alicecent} is fitted
(Fig. \ref{fig:ancestors}) with three different parametrizations of
the ancestor dependence mentioned above:
\begin{itemize}
\item a two-components model:             $\dNdeta \propto f\cdot\Npart+(1-f)\cdot\Ncoll$;
\item a power-law function of $\Npart$: $\dNdeta \propto \Npart^{\alpha}$;
\item a power-law function of $\Ncoll$: $\dNdeta \propto \Ncoll^{\beta}$.
\end{itemize}
and the fit parameters are reported in Tab. \ref{tab:FitPar}.  We note
that the value obtained for $f$ is in a good agreement with the value
obtained in the NBD-Glauber fit, shown in Fig.~\ref{fig:glau}.

While the value obtained for $\alpha$ and for $\beta$ with the
power-law parametrization of \Npart\ and \Ncoll\ indicate that neither
of these scalings perfectly describes the data ($\alpha >1$ and $\beta
<1$), we note that the value of $\alpha$ is similar to that measured
at RHIC (1.16 $\pm$ 0.04 \cite{Adler2005}) and slightly higher than
that at the SPS ($\alpha \sim 1$, see \cite{Abreu:2002fw} for a
review).  The results obtained with the two-component model, where
$0<f<1$, indicate that both the contribution of \Npart\ and \Ncoll\ are
needed to explain the particle production confirm this.  However, the
$\chi^2$/NDF reported in Table \ref{tab:FitPar}, indicate an equally
good fit for all models, thus revealing that no unique physics
conclusion can be drawn from such fits and that the particular choice
of parametrization has no influence on the results of the centrality
determination.

\begin{figure}[ht]
 \centering
 \includegraphics[width=0.75\textwidth]{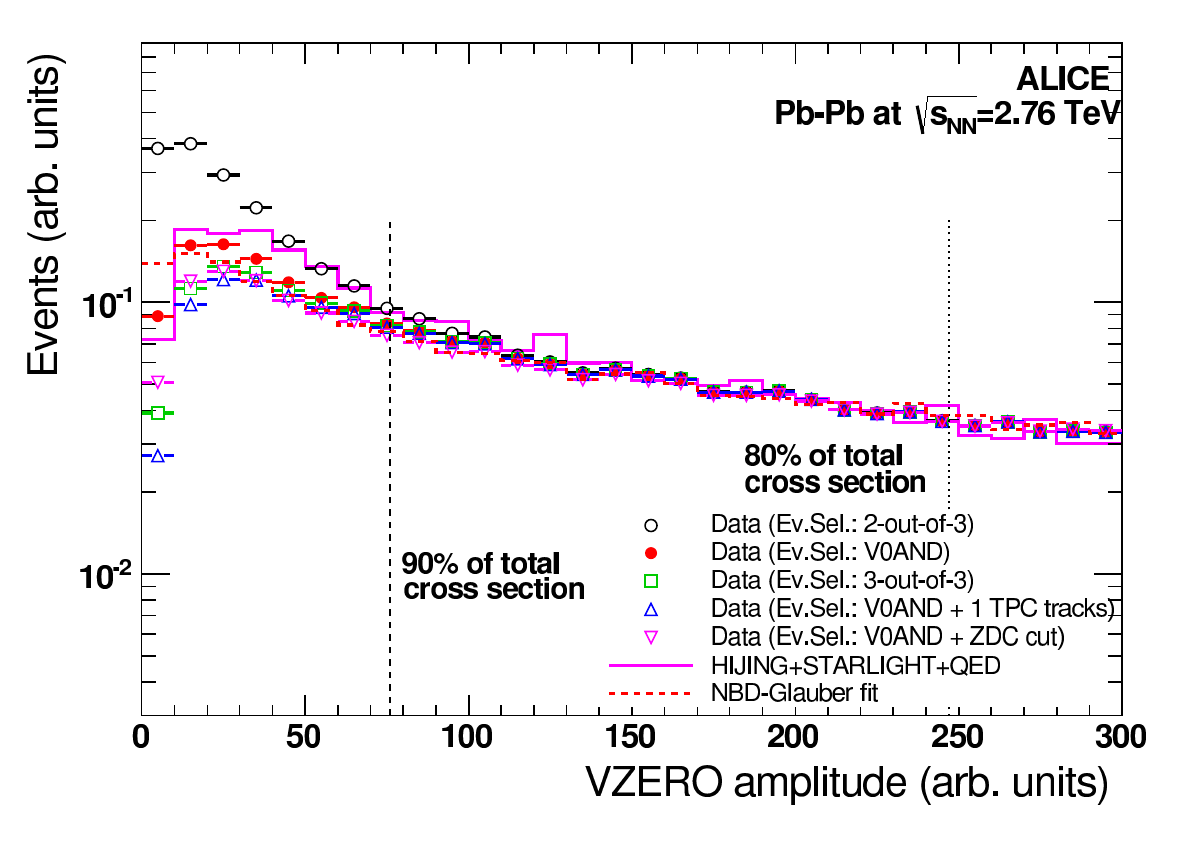}
 \caption{(Color online) Distribution of the \VZERO\ amplitude zoomed in the most
  peripheral region. The distribution is compared to the NBD-Glauber
  fit and to the sum of the HIJING + STARLIGHT + QED simulations.
  \label{fig:glauHijng}}
\end{figure}

\subsection{Systematic uncertainty on the \emph{Anchor Point}}
\label{subsec:sysAP}

The determination of the AP by either correcting or fitting
the multiplicity distribution is evaluated in Fig.~\ref{fig:glauHijng}
by comparing the \VZERO\ amplitude distributions for various event 
selections.  
The systematic uncertainty on the AP is estimated by comparing the
percentage of the hadronic cross section at the \VZERO\ amplitude
chosen as the AP (V0$_{\rm AP}$) obtained correcting or fitting
the multiplicity distribution.  For the first method
(Sec.~\ref{subsec:correctmult}), we used the results from the HIJING
and AMPT simulations. For the second method
(Sec.~\ref{subsec:NBD-glauber}), we used alternative centrality
definitions based on (i) \TPC\ tracks; (ii) \SPD\ hit multiplicities,
and obtaining a value for the V0$_{\rm AP}$ using the
correlation between \SPD\ or \TPC\ and \VZERO; (iii) different ranges
for the Glauber model fit; (iv) different ancestor dependence of the
particle production model to a power law of \Npart; (v) different
nucleon-nucleon cross section and parameters of the Woods-Saxon
distribution within their estimated uncertainties. All the results,
compared in Table \ref{tab:xsec}, allow to define the AP as the
\VZERO\ amplitude above which we obtain 90\% of the hadronic cross
section with the NBD-Glauber fit (the baseline in Tab.\ref{tab:xsec})
with a systematic uncertainty of 1\%, determined as the RMS of all the
results presented in Table \ref{tab:xsec}.  The variations of the AP
are not part of the quoted systematic uncertainties for \Npart,
\Ncoll\ and \TAB, which include only variations of the Glauber
parameters.  The uncertainty on the AP is typically included in our
analyses as an uncertainty on the limits of the centrality classes and
propagated into an uncertainty on the specific measured observable.

\begin{table*}[t]
 \caption{Comparison of the percentage of the hadronic cross section above the \VZERO\
 amplitude chosen as AP (V0$_{\rm AP}$) for various cases considered in the systematic 
 studies of the Glauber fits and with HIJING and AMPT simulations.
 \label{tab:xsec}}
\begin{tabular}{cc}
Method & \% of total cross section above the V0$_{\rm AP}$  \\
\hline
Glauber Fits &\\
Baseline & 90.00\\ 
(i) fit \TPC\ tracks & 89.88\\
(ii) fit \SPD\ clusters & 89.87\\
(iii) fit only 50\% of cross sec & 90.11\\
(iv) different ancestor dependence & 90.66\\
(v) different Wood-Saxons par & 90.43\\
\hline
HIJING simulations &\\
$2\mathrm{-out-of-}3$ & 92.50 \\
V0AND & 89.05 \\
$3\mathrm{-out-of-}3$ & 90.15 \\
V0AND + \TPC\ & 91.12 \\
V0AND + \ZDC\ & 89.52 \\
\hline
AMPT simulations & \\
$2\mathrm{-out-of-}3$   & 92.49 \\
V0AND        & 89.49 \\
$3\mathrm{-out-of-}3$  & 90.59 \\
V0AND + \TPC\  & 91.36 \\
V0AND + \ZDC\  & 89.00 \\
\hline
\end{tabular}
\end{table*}

\section{Centrality classes and their relation with geometrical quantities}
\label{sec:zdc}

\subsection{Determination of the centrality classes with the multiplicity distributions} 
\label{subsec:CentMult}

The percentile of the hadronic cross section is determined for any
value of the \VZERO\ amplitude by integrating the measured \VZERO\
amplitude distribution normalized at the anchor point V0$_{AP}$,
i.e.\ 90\% of the hadronic cross section. For example, if we define
$V$ as the \VZERO\ amplitude, the top 10\% central class is defined
by the boundary V0$_{10}$ which satisfies
\begin{equation}
\frac{\int_{V0_{\rm 10}}^{\infty} (dN_{\rm evt} / dV) \, dV}
{\int_{V0_{\rm AP}}^{\infty}      (dN_{\rm evt} / dV) \, dV }
= \frac{1}{9}
\end{equation}
The same is done for the number of clusters in the \SPD\ and the number of
reconstructed tracks in the \TPC. The events with multiplicity lower
than that of the anchor point, contaminated by EM background and
trigger inefficiency, are not used in the physics analyses.

One can divide the experimental distribution into classes, by defining
sharp cuts on e.g. \VZERO\ amplitude, which correspond to well defined
percentile intervals of the hadronic cross section. The number of
centrality classes that one can define is connected with the
resolution achieved on the quantities used in the definition. In
general, centrality classes are defined so that the separation between
the central values of $b$ and $\Npart$ for two adjacent classes is
significantly larger than the resolution of that variable (see
Sec. \ref{sec:resolution}).

\subsection{Finding the number of participants with the multiplicity distributions} 
\label{subsec:Npartdata}

In Sec.~\ref{subsec:NBD-glauber} we fit the measured \VZERO\ amplitude
distribution with the amplitude distribution simulated with the
NBD-Glauber.  This creates a connection between an experimental
observable and the geometrical model of nuclear collisions used in the
Glauber Monte Carlo.  From this we can access the geometrical
properties, like \Npart, \Ncoll, \TAB.
A given centrality class, defined by sharp cuts in the measured
distribution, corresponds to the same class in the simulated
distribution. For the simulated distribution we retain the input
information from the Glauber model. Therefore, we can calculate the
mean number of participants \avNpart, the mean number of collisions
\avNcoll\, and the average nuclear overlap function \avTAB\ for centrality
classes defined by sharp cuts in the simulated multiplicity
distribution, corresponding to given percentiles of the hadronic
cross section.
As shown in Table \ref{tab:Npartcompare}, the mean values and their
dispersions differ from those calculated for geometrical classes,
defined by sharp cuts in the impact parameter $b$
(Table~\ref{tab:Npart}), by less than 1\% for the most central classes
(up to about 50\%) and by less than 2\% for the most peripheral ones
(above 50\%). This confirms that multiplicity fluctuations and
detector resolution only play a minor role in the centrality
determination.

\subsection{Determination of the centrality classes with the ZDC} 
Another way to determine the centrality is to measure the energy
deposited by the spectators in the \ZDC.  The spectator neutrons and
protons having a rapidity close to that of the beam, are detected
in the \ZDC.
Naively, measurement of the number of spectator
neutrons and protons would give direct measurement of the number of
participants since \Npart\ would simply be given by 
\begin{equation} \label{eq:Npart} 
\ensuremath{N_\mathrm{part}} = 2  \mathrm A - E_{\mathrm ZDC}/ E_{\mathrm A}
\end{equation} 
where $E_{\mathrm ZDC}$ is the energy measured in the \ZDC, A = 208
is the mass number of Pb, and $E_{\mathrm A}$ is the beam energy per
nucleon. However, fragment formation amongst the spectator nucleons
breaks the simple linear and monotonic relation
in the measured variables, since some spectator nucleons are bound
into light nuclear fragments that have a charge over mass ratio
similar to the beam, therefore, remaining inside the beam-pipe and
are undetected by the \ZDC\ \cite{zdcTDR, zdcOppedisano}.  This
effect becomes quantitatively important for peripheral events and
therefore Eq.~\ref{eq:Npart} cannot be used as a reliable estimate of \Npart.

Consequently, the \ZDC\ information needs to be correlated to another
quantity that has a monotonic relation with \Npart. In our case, we
use the energy measured by two small EM calorimeters (\ZEM). These
detectors are placed only on the A side about 7.5 m from the
interaction point, covering the region $4.8 < \eta < 5.7$ 
\cite{aliceapp}.

Since the \ZDC\ calorimeters are far from the interaction region, and
therefore have an acceptance insensitive to the vertex position, a
centrality measurement based on the \ZDC\ is particularly suited for
any analysis that does not require a vertex cut \cite{VZEROeta}.

\begin{figure}[btp]
 \centering
 \includegraphics[width=0.6\textwidth]{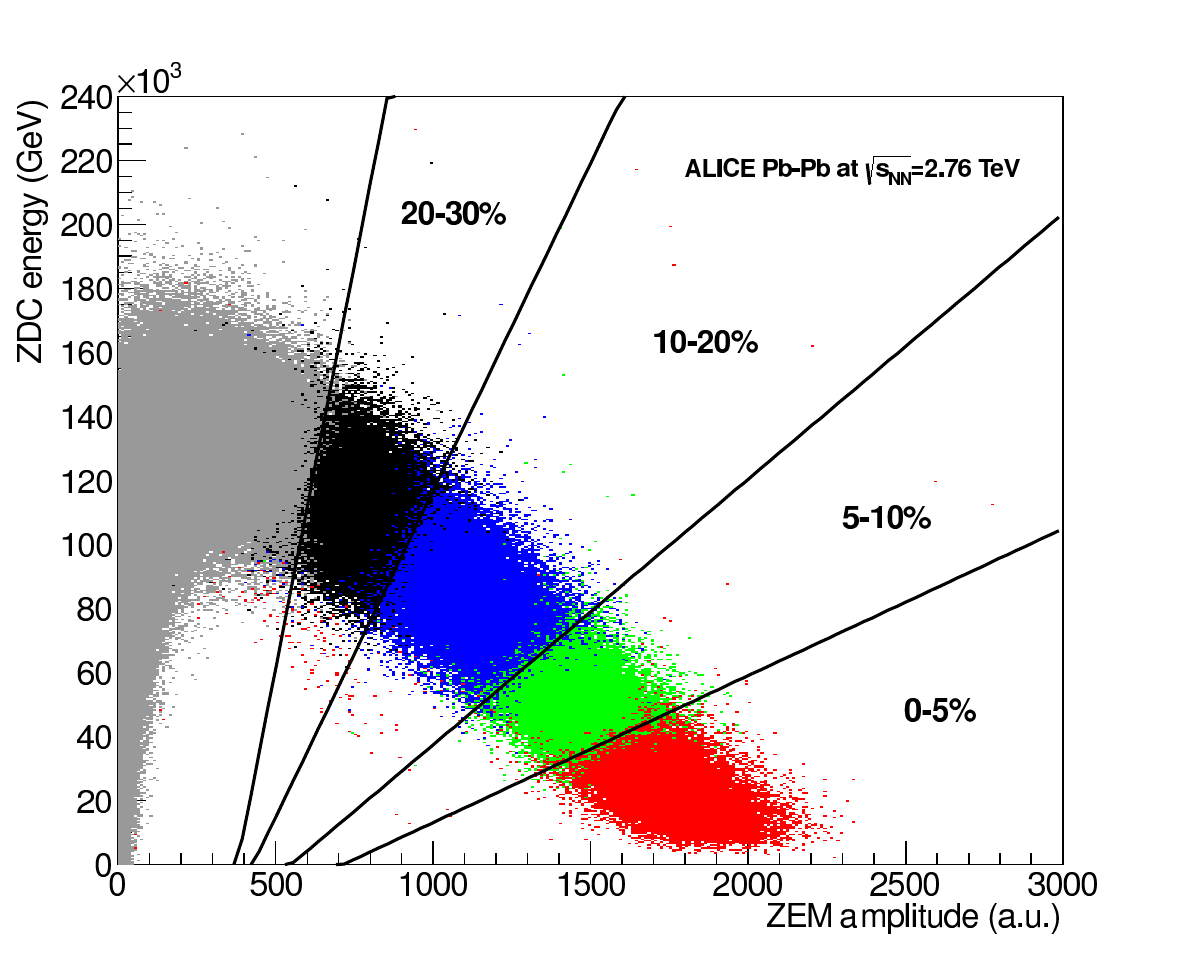}
 \caption{(Color online) Spectator energy deposited in the \ZDC\ calorimeters as a
  function of \ZEM\ amplitude. The same correlation is shown for
  different centrality classes (5\%, 10\%, 20\% and 30\%) obtained by
  selecting specific \VZERO\ amplitudes. The lines are a fit to the
  boundaries of the centrality classes with linear functions, where
  only the slope is fitted and the offset point is fixed (see text).  
  \label{fig:zdczem}}
\end{figure}

Centrality classes are defined by cuts on the two-dimensional
distribution of the \ZDC\ energy as a function of the \ZEM\ amplitude.
The \ZDC\ signal is proportional to \Npart\ for central events, while
the \ZEM\ amplitude is an unknown function of \Npart\ and \Ncoll.
Therefore the definition of the centrality classes in this
two-dimensional space is not trivial.  As shown in
Fig.~\ref{fig:zdczem}, centrality classes are defined by using the
centrality classes defined previously with the \VZERO\ amplitude to
determine regions in the \ZDC-\ZEM\ plane, corresponding to a given
centrality.  The boundaries between centrality classes, or the points
belonging to the same narrow centrality class $c$ ($c \pm \delta c$)
can be fitted with linear functions. All these lines are found to
intersect at a common point. Using this common point, we refitted the
boundaries of the various centrality classes with the linear functions
shown in Fig.~\ref{fig:zdczem}.

As can be seen from the figure, the slopes of the fitted functions
increase going from central to peripheral collisions and tend to
infinity, as the lines become almost straight vertical lines, when
approaching the point where the correlation between \ZDC\ and \ZEM\
inverts its sign.  The value of the slope that defines a centrality
class in the \ZDC\ vs \ZEM\ phase-space is proportional to the tangent
of the percentile, which implies that the percentiles behave like an
angle in the \ZDC\ vs \ZEM\ phase-space.

\begin{figure}[btp]
\centering
 \includegraphics[width=0.6\textwidth]{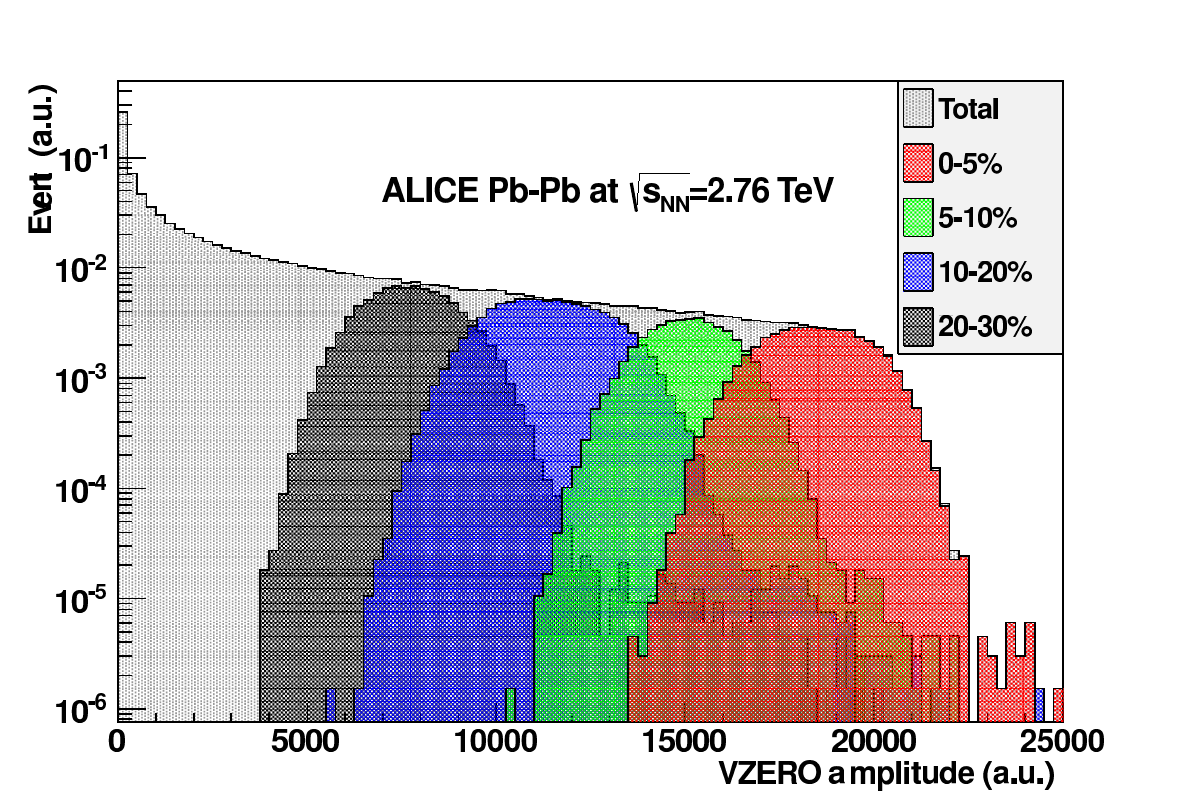}
 \caption{(Color online) \VZERO\ amplitude distribution of events of various
  centrality classes selected from the correlation between \ZDC\ and
  \ZEM\ amplitudes explained in the text.  
  \label{fig:v0zdcsel}}
\end{figure}

This function of \ZDC\ and \ZEM\ can then be used as centrality estimator
for the most central events (0--30\%) above the turning-point of \ZDC.
Figure \ref{fig:v0zdcsel} shows the distribution of the \VZERO\
amplitude for all triggered events and for various centrality classes
selected with this method. 

\section{Resolution of the centrality determination}
\label{sec:resolution}

As described above, two independent methods are used to determine
experimentally the centrality of the collision.  The first one uses
the multiplicity distributions from various detectors covering
different pseudo-rapidity ranges. Specifically we use the sum of the
amplitude in the \VZERO\ detectors (A and C side), the number of
clusters in the outer layer of the \SPD\ detector, and the number of
tracks reconstructed in the \TPC.  The second method uses the \ZDC\
correlated with the \ZEM.

\begin{figure}[ht]
 \centering
 \includegraphics[width=0.6\textwidth]{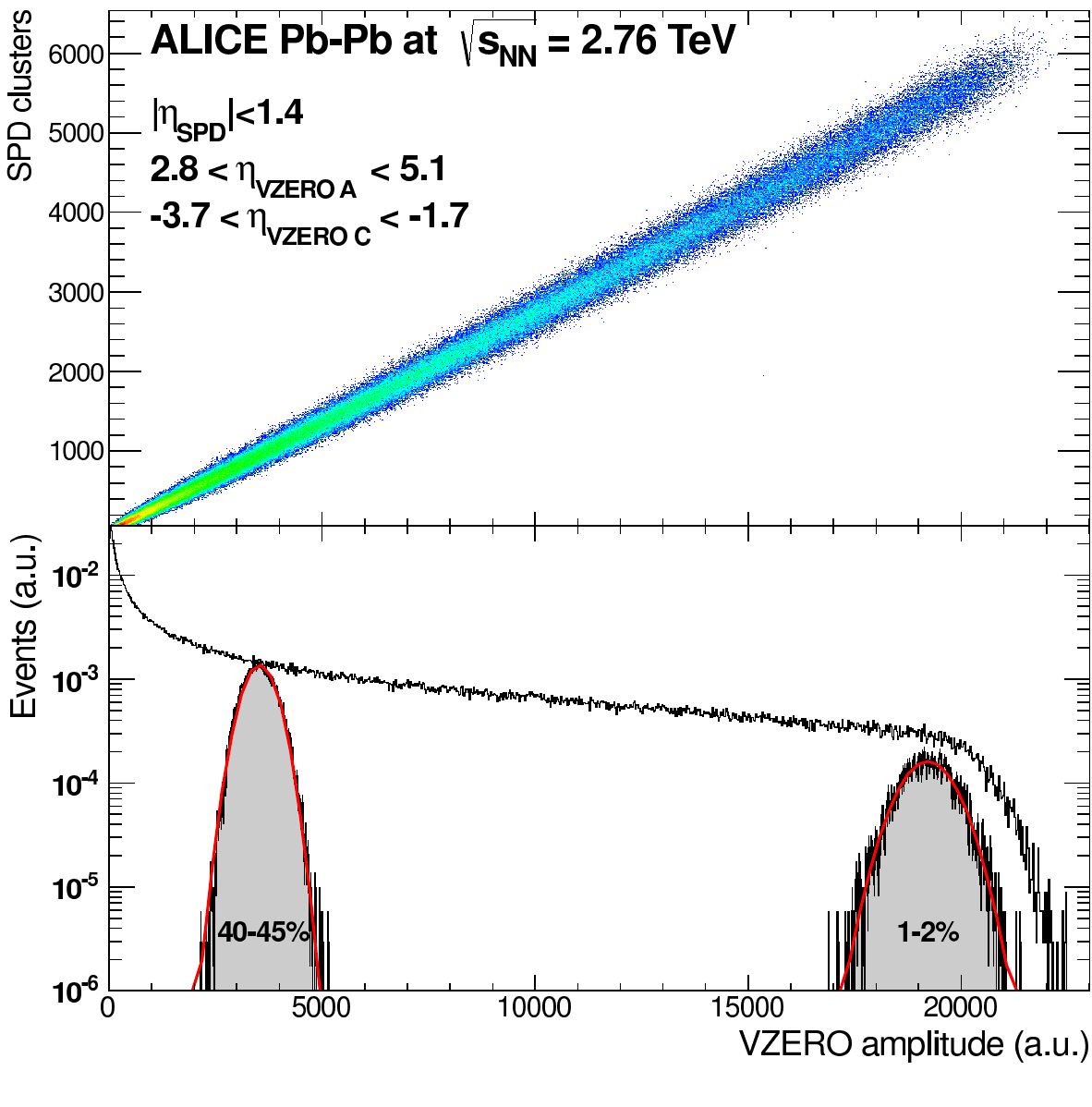}
 \caption{(Color online) Top: Correlation between \SPD\ multiplicity and \VZERO\
  amplitude. The rapidity coverage of each detector is
  indicated on the figure. Bottom: \VZERO\ amplitude
  distributions for the centrality classes selected by \SPD. Two
  centrality classes (1-2\% and 40-45\%) are indicated and fitted
  with a Gaussian.
  \label{fig:v0spd}}
\end{figure}

The accuracy of the experimental determination of the centrality was
evaluated by comparing the different estimates event-by-event.  For example,
in \Fig{fig:v0spd} we compare the estimates based on the \SPD\
multiplicity and the \VZERO\ amplitude.  The \VZERO\ amplitude
distribution is shown for two centrality classes selected by the
\SPD\ multiplicity. The distributions for the two centrality classes
are reasonably well fitted with a Gaussian distribution.

The resolution in the experimental definition of the centrality
classes is evaluated event-by-event as the RMS of the distribution of
the differences between the centrality determined over all estimators
and the mean value of the centrality for the event.
First the average value of the centrality $\langle c \rangle$ is
calculated for each event by averaging the centrality determined by
each estimator:
\begin{equation} \label{eq:CentRes0}
\langle c \rangle = \frac{\sum_{i=0}^{N} c_i }{N}.
\end{equation}
$c_i$ is the centrality of an event determined by an estimator $i$,
where $i$ is the index running over all N = 6 centrality estimators
used: \VZERO\ (A and C), \SPD, \TPC, \ZDC.
In the next step, the centrality is weighted by $\Delta_i = c_i -
\langle c \rangle$: the difference between the centrality
determined by each estimator and the mean value of the centrality from
Eq. \ref{eq:CentRes0}:
\begin{equation}
\langle c \rangle = \frac{\sum_{i=0}^{N} c_i / \Delta_i^2}{\sum_{i=0}^{N} 1 / \Delta_i^2}.
\end{equation}
This latter calculation is performed iteratively replacing $\langle c
\rangle$ by the new value until convergence is achieved which
typically occurs after the second iteration.
Finally, the centrality resolution of an estimator is
evaluated as the RMS of its $\Delta_i$ distribution for each
centrality.

\begin{figure}[tp]
 \centering
 \includegraphics[width=0.49\textwidth]{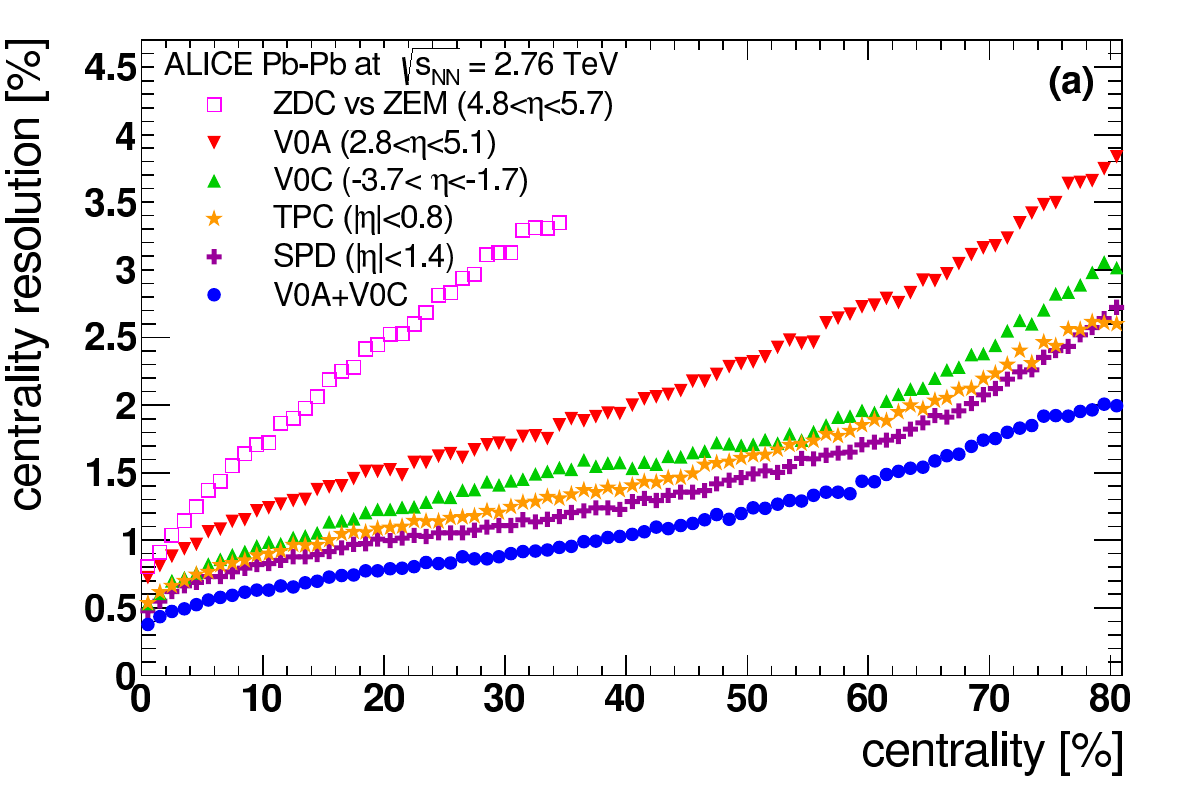}
 \includegraphics[width=0.49\textwidth]{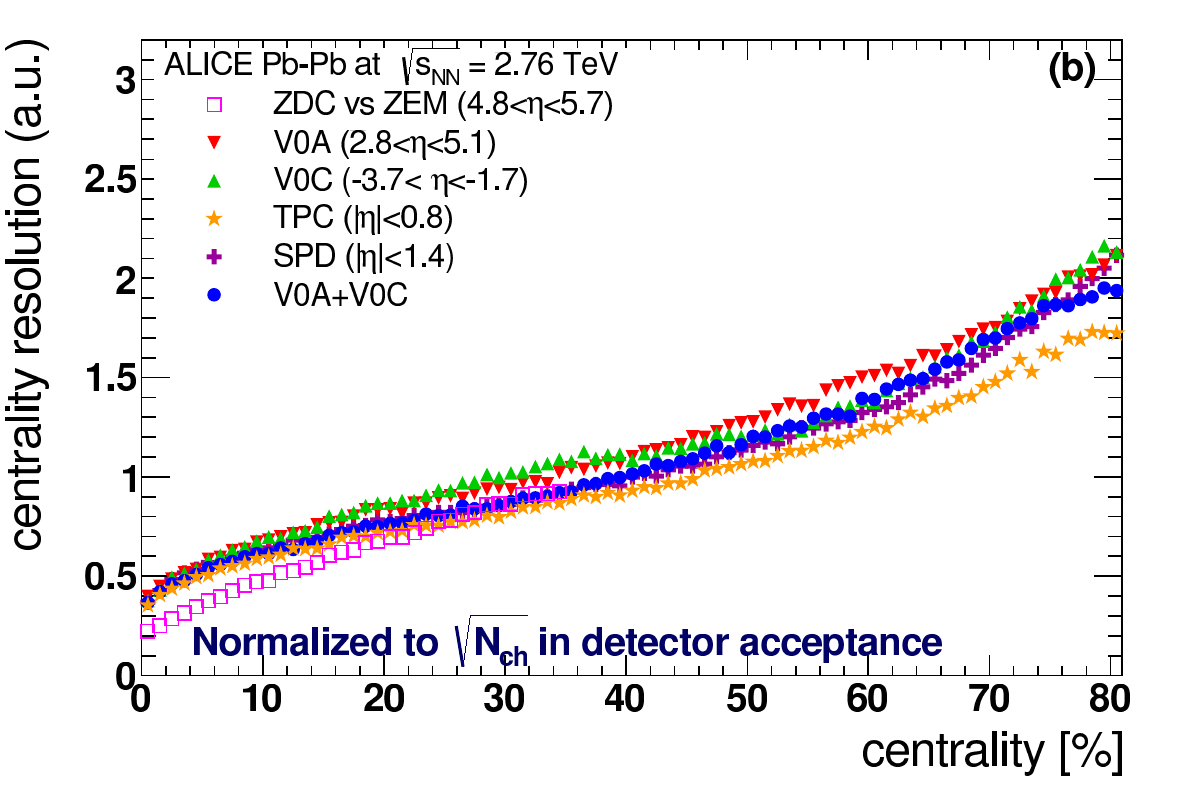}
 \caption{(Color online) Left: Centrality resolution $\Delta_i$ for all the
  estimators evaluated in the analysis. Right: Resolution, in
  arbitrary units, scaled by $\sqrt{N_\mathrm{ch}}$ measured in each
  detector.
  \label{fig:resolution}}
\end{figure}

The \ZDC-\ZEM\ estimator is ignored for peripheral events ($\langle c
\rangle > 35\% $) since its results are reliable only for the most central
collisions.  The resolution is shown in Fig.~\ref{fig:resolution}
(left panel) as a function of the centrality percentile.

The resolution depends on the rapidity coverage of the detector
used. The best centrality resolution is achieved when combining the
\VZEROA\ and \VZEROC\ detector, due to the large pseudo-rapidity
coverage (4.3 units in total). It ranges from 0.5\% in central to 2\%
in peripheral collisions. The resolution obtained with the \SPD\ and
the \TPC\ ranges from 1\% in central to 3\% in peripheral collisions
($\gtrsim 80\%$).

We measured the pseudo-rapidity dependence of the
charged particle multiplicity at midrapidity~\cite{alicemult} with
the \SPD, and at forward rapidity \cite{Toia2011} using all the
rapidity coverage of the \SPD, the \VZERO\ and the FMD detectors. The
total charged particle multiplicity $N_\mathrm{ch}$ is obtained by
integration. The centrality resolution was scaled by
$\sqrt{N_\mathrm{ch}}$ measured in the rapidity window of each
detector (see right panel of Fig.~\ref{fig:resolution}). The figure
shows that all the results are consistent on an arbitrary unit scale,
except for the \ZDC-\ZEM\ estimator which is better for central
collisions because it uses information from two detectors.

The centrality resolution was tested with a full HIJING and GEANT 
detector simulation.
In the HIJING simulations the true value of the event centrality
($c_{\rm true}$) is known for every given event.  After using GEANT one
obtains the signals in \VZERO, \SPD, \TPC\ for the given event and
hence using these centrality estimators can calculate the value of the
$\langle c \rangle$ for the given event with Eq.\ref{eq:CentRes0}.
The real centrality resolution, given for the given event by the
difference between the $c_{true}$ and the $\langle c \rangle$
calculated for each extimator, is consistent with the one calculated
with data.

\section{Summary}
\label{sec:conclusions}

Heavy-ion collisions can be characterized by the number of charged
particles produced in the collision. In principle, when normalized to
the trigger efficiency used to collect the data sample, the charged
particle multiplicity could provide a measurement of the hadronic
cross section. However, at the LHC the large cross section for EM 
processes contaminates the very peripheral collisions. 
This problem was overcome in two ways.

In the first method, dedicated simulations of hadronic and EM
processes (Fig.~\ref{fig:v0m_datasim}) were performed and data were
corrected for efficiency of the event selection
(Fig.~\ref{fig:trieff}) and purity of the event sample
(Fig.~\ref{fig:purity}).  In the second method, the measured
multiplicity distribution was fitted with a Glauber calculation
(Fig.~\ref{fig:glau}).  Both methods allow to determine a centrality
value above which the background contamination is negligible and the
event selection is fully efficient. The corresponding value of
multiplicity and centrality is defined as the anchor point, and is
used for the centrality normalization (Table~\ref{tab:xsec}).

Using the AP, the measured event sample can be divided in centrality
classes which correspond to well defined percentiles of the hadronic
cross section. Several approaches were developed.  The first method
uses charged particle multiplicity (measured by various detectors,
with different rapidity coverage, such as the \VZERO, the \SPD, and
the \TPC).  The second method uses the \ZDC, which measures the
nucleon spectators directly, as well as the correlation to the \ZEM\
energy in order to resolve the ambiguity due to nuclear fragmentation.
The centrality is obtained from linear functions that fit the contours
of the classes defined by the \VZERO, in the \ZDC-\ZEM\ plane
(Fig.~\ref{fig:zdczem}).  As standard method, typically used in ALICE
physics analyses, we used the NBD-Glauber fit to the \VZERO\ amplitude
(Fig.~\ref{fig:glau}) to determine the AP, and the other methods
described to asses a systematic uncertainty on the centrality
determination.

The resolution of the centrality determination, which depends on the
pseudo-rapidity coverage of the detector used, was determined as the
weighed RMS of all the estimates; it ranges from 0.5\% in central
to 2\% in peripheral collisions (Fig.~\ref{fig:resolution}).

Finally, mean numbers of the relevant geometrical quantities, such as
 \Npart\ and \Ncoll, were calculated for typical centrality classes,
 using the Glauber Model and the fit to the measured
 multiplicity distribution (Table ~\ref{tab:Npart}). This fit creates a
 mapping between a measured quantity and one obtained with a
 phenomenological calculation for which the geometrical properties are
 known.  The results, nearly identical to those obtained for
 centrality classes defined by classifying the events according to
 their impact parameter, provide a general tool to compare ALICE
 measurements with those of other experiments, at different energies
 and with different colliding systems as well as theoretical
 calculations.

\ifpreprint
\iffull
\newenvironment{acknowledgement}{\relax}{\relax}
\begin{acknowledgement}
\section*{Acknowledgements}
The ALICE collaboration would like to thank all its engineers and technicians for their invaluable contributions to the construction of the experiment and the CERN accelerator teams for the outstanding performance of the LHC complex.
\\
The ALICE collaboration acknowledges the following funding agencies for their support in building and
running the ALICE detector:
 \\
State Committee of Science, Calouste Gulbenkian Foundation from
Lisbon and Swiss Fonds Kidagan, Armenia;
 \\
Conselho Nacional de Desenvolvimento Cient\'{\i}fico e Tecnol\'{o}gico (CNPq), Financiadora de Estudos e Projetos (FINEP),
Funda\c{c}\~{a}o de Amparo \`{a} Pesquisa do Estado de S\~{a}o Paulo (FAPESP);
 \\
National Natural Science Foundation of China (NSFC), the Chinese Ministry of Education (CMOE)
and the Ministry of Science and Technology of China (MSTC);
 \\
Ministry of Education and Youth of the Czech Republic;
 \\
Danish Natural Science Research Council, the Carlsberg Foundation and the Danish National Research Foundation;
 \\
The European Research Council under the European Community's Seventh Framework Programme;
 \\
Helsinki Institute of Physics and the Academy of Finland;
 \\
French CNRS-IN2P3, the `Region Pays de Loire', `Region Alsace', `Region Auvergne' and CEA, France;
 \\
German BMBF and the Helmholtz Association;
\\
General Secretariat for Research and Technology, Ministry of
Development, Greece;
\\
Hungarian OTKA and National Office for Research and Technology (NKTH);
 \\
Department of Atomic Energy and Department of Science and Technology of the Government of India;
 \\
Istituto Nazionale di Fisica Nucleare (INFN) and Centro Fermi -
Museo Storico della Fisica e Centro Studi e Ricerche "Enrico
Fermi", Italy;
 \\
MEXT Grant-in-Aid for Specially Promoted Research, Ja\-pan;
 \\
Joint Institute for Nuclear Research, Dubna;
 \\
National Research Foundation of Korea (NRF);
 \\
CONACYT, DGAPA, M\'{e}xico, ALFA-EC and the HELEN Program (High-Energy physics Latin-American--European Network);
 \\
Stichting voor Fundamenteel Onderzoek der Materie (FOM) and the Nederlandse Organisatie voor Wetenschappelijk Onderzoek (NWO), Netherlands;
 \\
Research Council of Norway (NFR);
 \\
Polish Ministry of Science and Higher Education;
 \\
National Authority for Scientific Research - NASR (Autoritatea Na\c{t}ional\u{a} pentru Cercetare \c{S}tiin\c{t}ific\u{a} - ANCS);
 \\
Ministry of Education and Science of Russian Federation,
International Science and Technology Center, Russian Academy of
Sciences, Russian Federal Agency of Atomic Energy, Russian Federal
Agency for Science and Innovations and CERN-INTAS;
 \\
Ministry of Education of Slovakia;
 \\
Department of Science and Technology, South Africa;
 \\
CIEMAT, EELA, Ministerio de Educaci\'{o}n y Ciencia of Spain, Xunta de Galicia (Conseller\'{\i}a de Educaci\'{o}n),
CEA\-DEN, Cubaenerg\'{\i}a, Cuba, and IAEA (International Atomic Energy Agency);
 \\
Swedish Research Council (VR) and Knut $\&$ Alice Wallenberg
Foundation (KAW);
 \\
Ukraine Ministry of Education and Science;
 \\
United Kingdom Science and Technology Facilities Council (STFC);
 \\
The United States Department of Energy, the United States National
Science Foundation, the State of Texas, and the State of Ohio.
\end{acknowledgement}
\ifbibtex
\bibliographystyle{\bibstname}
\bibliography{biblio}{}
\else

\fi
\appendix
\section{Tables}
\label{sec:tables}

As described in Sec. \ref{sec:glauber}, for the physics in ALICE
analyses the average values of \Npart, \Ncoll, or \TAB\ for
centrality classes defined by sharp cuts in the impact parameter
distributions are used. These are reported in Table \ref{tab:Npart}.
Therefore \avNpart, \avNcoll\, and \avTAB\ depend exclusively on the
nuclear geometrical parameters, and not on any measured quantity. Their
uncertainty is calculated by varying the parameters of the Glauber
calculations (i.e. the parameters of the Woods-Saxon and the hadronic
cross section \signn) by the known uncertainty.  We label the \avNpart\
calculated with this procedure as \avNpartgeo.

Another possibility, discussed in Sec.\ref{subsec:Npartdata}, is to
define the average values of \Npart, \Ncoll, or \TAB\ for centrality
classes by sharp cuts in the fitted multiplicity
distribution. Following this strategy, it is also possible to
incorporate in the uncertainty, besides the uncertainties related to
the Glauber calculation, those related to the measurement of the AP:
the experimental region which is actually being used for the physics
analyses, because it is free of background and the trigger efficiency
is known.  In this case, the AP can be varied by the uncertainty that
was estimated (90\% $\pm$ 1\%) and recalculate \avNpart, \avNcoll\ and
\avTAB\ with these variations.  The \avNpart\ calculated with this
procedure is labled as \avNpartdata. The variations for the AP are labelled 
$\ensuremath{\langle N_\mathrm{part}^{\rm data +} \rangle}$
and 
$\ensuremath{\langle N_\mathrm{part}^{\rm data -} \rangle}$
respectively.

In Table \ref{tab:Npartcompare} \avNpartgeo\ is compared to
\avNpartdata\ for various centrality classes. The default values of
\avNpartdata\ are compared to the values obtained by varying the AP.
The discrepancies $\Delta$ are calculated as
\begin{equation}
  \Delta = \frac{|\ensuremath{\langle N_\mathrm{part}^{\rm geo} \rangle} - \ensuremath{\langle N_\mathrm{part}^{\rm data} \rangle} |}
  {(\ensuremath{\langle N_\mathrm{part}^{\rm geo} \rangle} + \ensuremath{\langle N_\mathrm{part}^{\rm data} \rangle} ) }
\end{equation}
Same comparison is done in Tables \ref{tab:Ncollcompare} and
\ref{tab:Taacompare} for \avNcoll and \avTAB respectively.  Table
\ref{tab:Npartcompare2} gives the comparison for the three quantities
but for bigger centrality classes.

\begin{table*}[t]
\footnotesize
\centering
\caption{$\Npart$ for \PbPb\ collisions at $\snn$ =~2.76~TeV with the corresponding uncertainties derived from a Glauber
 calculation. The \avNpartdata\ are calculated from the NBD-Glauber fit to the \VZERO\ amplitude, while the \avNpartgeo\ 
 are obtained by slicing the impact parameter distribution. \avNpartdata\ is also calculated for two variations of the AP, 
 i.e.\ moving it to 91\% ($\ensuremath{\langle N_\mathrm{part}^{\rm data +} \rangle}$) and to 
 89\% ($\ensuremath{\langle N_\mathrm{part}^{\rm data +} \rangle}$) respectively. 
 The last three columns report the discrepancies between \avNpartgeo\ and \avNpartdata\ 
 and \avNpartdata with the uncertainty of the AP.
 \label{tab:Npartcompare}}
\begin{tabular}{c|c|cc|cc|ccc}
Cent.
& $\ensuremath{\langle N_\mathrm{part}^{\rm geo} \rangle}  $ (syst. \%) 
& $\ensuremath{\langle N_\mathrm{part}^{\rm data} \rangle} $ 
& RMS
& $\ensuremath{\langle N_\mathrm{part}^{\rm data +} \rangle}$
& $\ensuremath{\langle N_\mathrm{part}^{\rm data -} \rangle}$
& $\Delta_{geo}^{data}$ (\%)
& $\Delta_{data}^{data +}$ (\%)
& $\Delta_{data}^{data -}$ (\%) \\
\hline
 0-1\%    & 403.8 (0.35) & 400.8 & 7.8 & 400.8 & 400.7 & 0.38 & 0.0041 & 0.0066  \\
 1-2\%    & 393.6 (0.46) & 392.5 & 11 & 392.7 & 392.3 & 0.14 & 0.021 & 0.018  \\
 2-3\%    & 382.9 (0.6) & 382.9 & 12 & 383.2 & 382.6 & 0.00013 & 0.037 & 0.035  \\
 3-4\%    & 372 (0.73) & 372.2 & 13 & 372.7 & 371.8 & 0.033 & 0.057 & 0.062  \\
 4-5\%    & 361.1 (0.83) & 361.4 & 13 & 362 & 360.9 & 0.047 & 0.081 & 0.072  \\
 5-10\%   & 329.4 (1.1) & 329.7 & 20 & 330.5 & 328.8 & 0.047 & 0.12 & 0.13  \\
 10-15\%  & 281.2 (1.4) & 281.6 & 18 & 282.8 & 280.3 & 0.064 & 0.21 & 0.23  \\
 15-20\%  & 239 (1.6) & 239.5 & 17 & 241 & 237.9 & 0.099 & 0.31 & 0.32  \\
 20-25\%  & 202.1 (1.8) & 202.7 & 15 & 204.4 & 200.9 & 0.14 & 0.42 & 0.43  \\
 25-30\%  & 169.5 (1.9) & 170.1 & 14 & 171.9 & 168.2 & 0.17 & 0.53 & 0.56  \\
 30-35\%  & 141 (2) & 141.7 & 12 & 143.6 & 139.7 & 0.24 & 0.66 & 0.7  \\
 35-40\%  & 116 (2.2) & 116.7 & 11 & 118.6 & 114.7 & 0.31 & 0.81 & 0.86  \\
 40-45\%  & 94.11 (2.1) & 94.77 & 9.7 & 96.68 & 92.83 & 0.35 & 1 & 1  \\
 45-50\%  & 75.3 (2.3) & 75.91 & 8.4 & 77.72 & 74.02 & 0.4 & 1.2 & 1.3  \\
 50-55\%  & 59.24 (2.5) & 59.77 & 7.3 & 61.49 & 58.02 & 0.44 & 1.4 & 1.5  \\
 55-60\%  & 45.58 (2.9) & 46.1 & 6.3 & 47.66 & 44.47 & 0.57 & 1.7 & 1.8  \\
 60-65\%  & 34.33 (2.6) & 34.65 & 5.4 & 36.09 & 33.2 & 0.47 & 2 & 2.1  \\
 65-70\%  & 25.21 (4) & 25.38 & 4.5 & 26.62 & 24.16 & 0.34 & 2.4 & 2.5  \\
 70-75\%  & 17.96 (3.3) & 18.06 & 3.8 & 19.07 & 17 & 0.27 & 2.7 & 3  \\
 75-80\%  & 12.58 (3.7) & 12.45 & 3 & 13.25 & 11.61 & 0.54 & 3.1 & 3.5  \\
 80-85\%  & 8.812 (2.8) & 8.275 & 2.4 & 8.914 & 7.646 & 3.1 & 3.7 & 4  \\
 85-90\%  & 6.158 (2.4) & 5.516 & 1.8 & 6.035 & 3.406 & 5.5 & 4.5 & 24  \\
\end{tabular}      
\end{table*}

\begin{table*}[t]
\footnotesize
\centering
\caption{Same as Table \ref{tab:Npartcompare} for $\Ncoll$.
\label{tab:Ncollcompare}}
\begin{tabular}{c|c|cc|cc|ccc}
Cent.
& $\ensuremath{\langle N_\mathrm{coll}^{\rm geo} \rangle}  $ (syst. \%) 
& $\ensuremath{\langle N_\mathrm{coll}^{\rm data} \rangle} $ 
& RMS
& $\ensuremath{\langle N_\mathrm{coll}^{\rm data +} \rangle}$
& $\ensuremath{\langle N_\mathrm{coll}^{\rm data -} \rangle}$
& $\Delta_{\rm geo}^{\rm data}$ (\%)
& $\Delta_{\rm data}^{\rm data +}$ (\%)
& $\Delta_{\rm data}^{\rm data -}$ (\%) \\
\hline
 0-1\%    & 1861 (8.2) & 1863 & 83 & 1864 & 1863 & 0.059 & 0.016 & 0.018  \\
 1-2\%    & 1766 (8.2) & 1761 & 79 & 1762 & 1759 & 0.15 & 0.04 & 0.047  \\
 2-3\%    & 1678 (8.2) & 1678 & 79 & 1680 & 1676 & 0.0063 & 0.063 & 0.067  \\
 3-4\%    & 1597 (8.3) & 1596 & 78 & 1599 & 1592 & 0.039 & 0.096 & 0.1  \\
 4-5\%    & 1520 (8.1) & 1520 & 77 & 1524 & 1516 & 0.002 & 0.12 & 0.12  \\
 5-10\%   & 1316 (8.2) & 1316 & 110 & 1321 & 1310 & 0.0099 & 0.2 & 0.2  \\
 10-15\%  & 1032 (8.2) & 1034 & 95 & 1040 & 1027 & 0.083 & 0.32 & 0.34  \\
 15-20\%  & 809.8 (8) & 811.7 & 80 & 819.2 & 803.9 & 0.11 & 0.46 & 0.48  \\
 20-25\%  & 629.6 (7.8) & 631 & 68 & 639 & 622.9 & 0.11 & 0.63 & 0.65  \\
 25-30\%  & 483.7 (7.5) & 485.8 & 57 & 493.6 & 477.7 & 0.22 & 0.79 & 0.85  \\
 30-35\%  & 366.7 (7.4) & 368.4 & 49 & 375.9 & 360.7 & 0.23 & 1 & 1.1  \\
 35-40\%  & 273.4 (7.4) & 274.8 & 40 & 281.7 & 267.8 & 0.26 & 1.2 & 1.3  \\
 40-45\%  & 199.4 (6.9) & 200.7 & 33 & 206.8 & 194.5 & 0.32 & 1.5 & 1.6  \\
 45-50\%  & 143.1 (6.6) & 143.8 & 26 & 149 & 138.6 & 0.26 & 1.8 & 1.9  \\
 50-55\%  & 100.1 (6.5) & 100.6 & 20 & 104.9 & 96.22 & 0.25 & 2.1 & 2.2  \\
 55-60\%  & 68.46 (6.2) & 68.7 & 15 & 72.12 & 65.28 & 0.18 & 2.4 & 2.6  \\
 60-65\%  & 45.79 (5.7) & 45.79 & 12 & 48.47 & 43.12 & 0.0038 & 2.8 & 3  \\
 65-70\%  & 29.92 (6.8) & 29.66 & 8.3 & 31.68 & 27.73 & 0.43 & 3.3 & 3.4  \\
 70-75\%  & 19.08 (5.7) & 18.82 & 5.8 & 20.22 & 17.39 & 0.68 & 3.6 & 4  \\
 75-80\%  & 12.07 (5.7) & 11.62 & 4 & 12.61 & 10.63 & 1.9 & 4.1 & 4.4  \\
 80-85\%  & 7.682 (5.1) & 6.925 & 2.7 & 7.595 & 6.269 & 5.2 & 4.6 & 5  \\
 85-90\%  & 4.904 (4.1) & 4.148 & 1.8 & 4.651 & 2.257 & 8.4 & 5.7 & 30  \\
\end{tabular}      
\end{table*}

\begin{table*}[t]
\footnotesize
\centering
\caption{Same as Table \ref{tab:Npartcompare} for $\TAB$.
\label{tab:Taacompare}}
\begin{tabular}{c|c|cc|cc|ccc}
  Cent.
& $\ensuremath{\langle T_\mathrm{AB}^{\rm geo} \rangle}  $ (syst. \%) 
& $\ensuremath{\langle T_\mathrm{AB}^{\rm data} \rangle} $ 
& RMS
& $\ensuremath{\langle T_\mathrm{AB}^{\rm data +} \rangle}$
& $\ensuremath{\langle T_\mathrm{AB}^{\rm data -} \rangle}$
& $\Delta_{geo}^{data}$ (\%)
& $\Delta_{data}^{data +}$ (\%)
& $\Delta_{data}^{data -}$ (\%) \\
\hline
 0-1\%    & 29.08 (3.2) & 29.11 & 1.3 & 29.12 & 29.1 & 0.056 & 0.016 & 0.018  \\
 1-2\%    & 27.6 (3.2) & 27.51 & 1.2 & 27.53 & 27.48 & 0.16 & 0.04 & 0.047  \\
 2-3\%    & 26.22 (3.2) & 26.22 & 1.2 & 26.25 & 26.19 & 0.0038 & 0.063 & 0.067  \\
 3-4\%    & 24.95 (3.2) & 24.93 & 1.2 & 24.98 & 24.88 & 0.032 & 0.096 & 0.1  \\
 4-5\%    & 23.75 (3.2) & 23.75 & 1.2 & 23.8 & 23.69 & 0.0019 & 0.12 & 0.12  \\
 5-10\%   & 20.56 (3.3) & 20.56 & 1.8 & 20.64 & 20.47 & 0.0036 & 0.2 & 0.2  \\
 10-15\%  & 16.13 (3.6) & 16.15 & 1.5 & 16.26 & 16.04 & 0.068 & 0.32 & 0.34  \\
 15-20\%  & 12.65 (3.7) & 12.68 & 1.2 & 12.8 & 12.56 & 0.13 & 0.46 & 0.48  \\
 20-25\%  & 9.837 (3.7) & 9.86 & 1.1 & 9.984 & 9.733 & 0.12 & 0.63 & 0.65  \\
 25-30\%  & 7.558 (3.4) & 7.591 & 0.9 & 7.713 & 7.463 & 0.22 & 0.79 & 0.85  \\
 30-35\%  & 5.73 (3.3) & 5.756 & 0.76 & 5.873 & 5.636 & 0.23 & 1 & 1.1  \\
 35-40\%  & 4.272 (3.7) & 4.294 & 0.63 & 4.402 & 4.184 & 0.26 & 1.2 & 1.3  \\
 40-45\%  & 3.115 (3.9) & 3.136 & 0.51 & 3.231 & 3.039 & 0.33 & 1.5 & 1.6  \\
 45-50\%  & 2.235 (4.2) & 2.248 & 0.41 & 2.328 & 2.165 & 0.28 & 1.8 & 1.9  \\
 50-55\%  & 1.564 (4.7) & 1.572 & 0.32 & 1.639 & 1.504 & 0.25 & 2.1 & 2.2  \\
 55-60\%  & 1.07 (5.2) & 1.073 & 0.24 & 1.127 & 1.02 & 0.16 & 2.4 & 2.6  \\
 60-65\%  & 0.7154 (5) & 0.7154 & 0.18 & 0.7573 & 0.6737 & 0.0007 & 2.8 & 3  \\
 65-70\%  & 0.4674 (6.2) & 0.4635 & 0.13 & 0.4949 & 0.4333 & 0.42 & 3.3 & 3.4  \\
 70-75\%  & 0.2981 (6.4) & 0.2941 & 0.091 & 0.3159 & 0.2717 & 0.68 & 3.6 & 4  \\
 75-80\%  & 0.1885 (6.9) & 0.1815 & 0.062 & 0.197 & 0.1661 & 1.9 & 4.1 & 4.4  \\
 80-85\%  & 0.12 (6.5) & 0.1082 & 0.042 & 0.1187 & 0.09795 & 5.2 & 4.6 & 5  \\
 85-90\%  & 0.07662 (5.9) & 0.06481 & 0.028 & 0.07267 & 0.03526 & 8.4 & 5.7 & 30  \\
\end{tabular}
\end{table*}

\begin{table*}[t]
\footnotesize
\centering
\caption{Same as above with bigger centrality classes
\label{tab:Npartcompare2}}
\begin{tabular}{c|c|cc|cc|ccc}
Cent.
& $\ensuremath{\langle N_\mathrm{part}^{\rm geo} \rangle}  $ (syst. \%) 
& $\ensuremath{\langle N_\mathrm{part}^{\rm data} \rangle} $ 
& RMS
& $\ensuremath{\langle N_\mathrm{part}^{\rm data +} \rangle}$
& $\ensuremath{\langle N_\mathrm{part}^{\rm data -} \rangle}$
& $\Delta_{geo}^{data}$ (\%)
& $\Delta_{data}^{data +}$ (\%)
& $\Delta_{data}^{data -}$ (\%) \\
\hline
0-5\%   &382.7 (0.77) & 382 & 17 & 382.3 & 381.7 & 0.096 & 0.04 & 0.038  \\
5-10\%  &329.6 (1.3) & 329.7 & 18 & 330.5 & 328.8 & 0.015 & 0.13 & 0.13  \\
10-20\% &260.1 (1.5) & 260.5 & 27 & 261.9 & 259.1 & 0.079 & 0.26 & 0.27  \\
20-40\% &157.2 (2) & 157.8 & 35 & 159.6 & 155.9 & 0.19 & 0.58 & 0.6  \\
40-60\% &68.56 (2.9) & 69.13 & 22 & 70.89 & 67.35 & 0.42 & 1.3 & 1.3  \\
60-80\% &22.52 (3.4) & 22.64 & 12 & 23.76 & 21.51 & 0.27 & 2.4 & 2.6  \\
\hline             
& $\ensuremath{\langle N_\mathrm{coll}^{\rm geo} \rangle}  $ (syst. \%) 
& $\ensuremath{\langle N_\mathrm{coll}^{\rm data} \rangle} $ 
& RMS
& $\ensuremath{\langle N_\mathrm{coll}^{\rm data +} \rangle}$
& $\ensuremath{\langle N_\mathrm{coll}^{\rm data -} \rangle}$
& $\Delta_{geo}^{data}$ (\%)
& $\Delta_{data}^{data +}$ (\%)
& $\Delta_{data}^{data -}$ (\%) \\
\hline
0-5\%   &1685 (11) & 1684 & 1.4e+02 & 1686 & 1681 & 0.044 & 0.065 & 0.067  \\
5-10\%  &1316 (11) & 1316 & 1.1e+02 & 1321 & 1310 & 0.011 & 0.2 & 0.2  \\
10-20\% &921.2 (10) & 922.7 & 1.4e+02 & 929.8 & 915.3 & 0.079 & 0.39 & 0.4  \\
20-40\% &438.4 (9.7) & 440 & 1.5e+02 & 447.5 & 432.3 & 0.19 & 0.85 & 0.89  \\
40-60\% &127.7 (8.8) & 128.4 & 59 & 133.2 & 123.7 & 0.29 & 1.8 & 1.9  \\
60-80\% &26.71 (7.3) & 26.48 & 18 & 28.25 & 24.74 & 0.43 & 3.2 & 3.4  \\

\hline
& $\ensuremath{\langle T_\mathrm{AB}^{\rm geo} \rangle}  $ (syst. \%) 
& $\ensuremath{\langle T_\mathrm{AB}^{\rm data} \rangle} $ 
& RMS
& $\ensuremath{\langle T_\mathrm{AB}^{\rm data +} \rangle}$
& $\ensuremath{\langle T_\mathrm{AB}^{\rm data -} \rangle}$
& $\Delta_{geo}^{data}$ (\%)
& $\Delta_{data}^{data +}$ (\%)
& $\Delta_{data}^{data -}$ (\%) \\
\hline            
0-5\%   &26.32 (3.2) & 26.31 & 2.2 & 26.34 & 26.27 & 0.028 & 0.066 & 0.066  \\
5-10\%  &20.56 (3.3) & 20.56 & 1.7 & 20.64 & 20.47 & 0.0051 & 0.2 & 0.2  \\
10-20\% &14.39 (3.1) & 14.42 & 2.2 & 14.53 & 14.3 & 0.092 & 0.39 & 0.4  \\
20-40\% &6.85 (3.3) & 6.876 & 2.3 & 6.993 & 6.754 & 0.19 & 0.85 & 0.89  \\
40-60\% &1.996 (4.9) & 2.007 & 0.92 & 2.081 & 1.933 & 0.28 & 1.8 & 1.9  \\
60-80\% &0.4174 (6.3) & 0.4137 & 0.29 & 0.4414 & 0.3865 & 0.44 & 3.2 & 3.4  \\
\end{tabular}
\end{table*}

\clearpage
\section{ALICE Collaboration}
\label{app:collab}

\begingroup
\small
\begin{flushleft}
B.~Abelev\Irefn{org1234}\And
J.~Adam\Irefn{org1274}\And
D.~Adamov\'{a}\Irefn{org1283}\And
A.M.~Adare\Irefn{org1260}\And
M.M.~Aggarwal\Irefn{org1157}\And
G.~Aglieri~Rinella\Irefn{org1192}\And
M.~Agnello\Irefn{org1313}\textsuperscript{,}\Irefn{org1017688}\And
A.G.~Agocs\Irefn{org1143}\And
A.~Agostinelli\Irefn{org1132}\And
Z.~Ahammed\Irefn{org1225}
% ...
\renewcommand\labelenumi{\textsuperscript{\theenumi}~}
\section*{Affiliation notes}
\renewcommand\theenumi{\roman{enumi}}
\begin{Authlist}
\item \Adef{M.V.Lomonosov}Also at: M.V.Lomonosov Moscow State University, D.V.Skobeltsyn Institute of Nuclear Physics, Moscow, Russia
\item \Adef{University of Belgrade, Faculty of Physics and "Vinvca" Institute of Nuclear Sciences, Belgrade, Serbia}Also at: University of Belgrade, Faculty of Physics and "Vinvca" Institute of Nuclear Sciences, Belgrade, Serbia
\item \Adef{Institute of Theoretical Physics, University of Wroclaw, Wroclaw, Poland}Also at: Institute of Theoretical Physics, University of Wroclaw, Wroclaw, Poland
\end{Authlist}
\section*{Collaboration Institutes}
\renewcommand\theenumi{\arabic{enumi}~}
\begin{Authlist}
\item \Idef{org1332}A. I. Alikhanyan National Science Laboratory (Yerevan Physics Institute) Foundation, Yerevan, Armenia
\item \Idef{org1279}Benem\'{e}rita Universidad Aut\'{o}noma de Puebla, Puebla, Mexico
\item \Idef{org1220}Bogolyubov Institute for Theoretical Physics, Kiev, Ukraine
\item \Idef{org20959}Bose Institute, Department of Physics and Centre for Astroparticle Physics and Space Science (CAPSS), Kolkata, India
\item \Idef{org1262}Budker Institute for Nuclear Physics, Novosibirsk, Russia
\item \Idef{org1292}California Polytechnic State University, San Luis Obispo, California, United States
\item \Idef{org1329}Central China Normal University, Wuhan, China
\item \Idef{org14939}Centre de Calcul de l'IN2P3, Villeurbanne, France
\item \Idef{org1197}Centro de Aplicaciones Tecnol\'{o}gicas y Desarrollo Nuclear (CEADEN), Havana, Cuba
\item \Idef{org1242}Centro de Investigaciones Energ\'{e}ticas Medioambientales y Tecnol\'{o}gicas (CIEMAT), Madrid, Spain
\item \Idef{org1244}Centro de Investigaci\'{o}n y de Estudios Avanzados (CINVESTAV), Mexico City and M\'{e}rida, Mexico
\item \Idef{org1335}Centro Fermi - Museo Storico della Fisica e Centro Studi e Ricerche ``Enrico Fermi'', Rome, Italy
\item \Idef{org17347}Chicago State University, Chicago, United States
\item \Idef{org1288}Commissariat \`{a} l'Energie Atomique, IRFU, Saclay, France
\item \Idef{org15782}COMSATS Institute of Information Technology (CIIT), Islamabad, Pakistan
\item \Idef{org1294}Departamento de F\'{\i}sica de Part\'{\i}culas and IGFAE, Universidad de Santiago de Compostela, Santiago de Compostela, Spain
\item \Idef{org1106}Department of Physics Aligarh Muslim University, Aligarh, India
\item \Idef{org1121}Department of Physics and Technology, University of Bergen, Bergen, Norway
\item \Idef{org1162}Department of Physics, Ohio State University, Columbus, Ohio, United States
\item \Idef{org1300}Department of Physics, Sejong University, Seoul, South Korea
\item \Idef{org1268}Department of Physics, University of Oslo, Oslo, Norway
\item \Idef{org1312}Dipartimento di Fisica dell'Universit\`{a} and Sezione INFN, Turin, Italy
\item \Idef{org1145}Dipartimento di Fisica dell'Universit\`{a} and Sezione INFN, Cagliari, Italy
\item \Idef{org1315}Dipartimento di Fisica dell'Universit\`{a} and Sezione INFN, Trieste, Italy
\item \Idef{org1285}Dipartimento di Fisica dell'Universit\`{a} `La Sapienza' and Sezione INFN, Rome, Italy
\item \Idef{org1154}Dipartimento di Fisica e Astronomia dell'Universit\`{a} and Sezione INFN, Catania, Italy
\item \Idef{org1132}Dipartimento di Fisica e Astronomia dell'Universit\`{a} and Sezione INFN, Bologna, Italy
\item \Idef{org1270}Dipartimento di Fisica e Astronomia dell'Universit\`{a} and Sezione INFN, Padova, Italy
\item \Idef{org1290}Dipartimento di Fisica `E.R.~Caianiello' dell'Universit\`{a} and Gruppo Collegato INFN, Salerno, Italy
\item \Idef{org1103}Dipartimento di Scienze e Innovazione Tecnologica dell'Universit\`{a} del Piemonte Orientale and Gruppo Collegato INFN, Alessandria, Italy
\item \Idef{org1114}Dipartimento Interateneo di Fisica `M.~Merlin' and Sezione INFN, Bari, Italy
\item \Idef{org1237}Division of Experimental High Energy Physics, University of Lund, Lund, Sweden
\item \Idef{org1192}European Organization for Nuclear Research (CERN), Geneva, Switzerland
\item \Idef{org1227}Fachhochschule K\"{o}ln, K\"{o}ln, Germany
\item \Idef{org1122}Faculty of Engineering, Bergen University College, Bergen, Norway
\item \Idef{org1136}Faculty of Mathematics, Physics and Informatics, Comenius University, Bratislava, Slovakia
\item \Idef{org1274}Faculty of Nuclear Sciences and Physical Engineering, Czech Technical University in Prague, Prague, Czech Republic
\item \Idef{org1229}Faculty of Science, P.J.~\v{S}af\'{a}rik University, Ko\v{s}ice, Slovakia
\item \Idef{org1184}Frankfurt Institute for Advanced Studies, Johann Wolfgang Goethe-Universit\"{a}t Frankfurt, Frankfurt, Germany
\item \Idef{org1215}Gangneung-Wonju National University, Gangneung, South Korea
\item \Idef{org20958}Gauhati University, Department of Physics, Guwahati, India
\item \Idef{org1212}Helsinki Institute of Physics (HIP) and University of Jyv\"{a}skyl\"{a}, Jyv\"{a}skyl\"{a}, Finland
\item \Idef{org1203}Hiroshima University, Hiroshima, Japan
\item \Idef{org1254}Indian Institute of Technology Bombay (IIT), Mumbai, India
\item \Idef{org36378}Indian Institute of Technology Indore, Indore, India (IITI)
\item \Idef{org1266}Institut de Physique Nucl\'{e}aire d'Orsay (IPNO), Universit\'{e} Paris-Sud, CNRS-IN2P3, Orsay, France
\item \Idef{org1277}Institute for High Energy Physics, Protvino, Russia
\item \Idef{org1249}Institute for Nuclear Research, Academy of Sciences, Moscow, Russia
\item \Idef{org1320}Nikhef, National Institute for Subatomic Physics and Institute for Subatomic Physics of Utrecht University, Utrecht, Netherlands
\item \Idef{org1250}Institute for Theoretical and Experimental Physics, Moscow, Russia
\item \Idef{org1230}Institute of Experimental Physics, Slovak Academy of Sciences, Ko\v{s}ice, Slovakia
\item \Idef{org1127}Institute of Physics, Bhubaneswar, India
\item \Idef{org1275}Institute of Physics, Academy of Sciences of the Czech Republic, Prague, Czech Republic
\item \Idef{org1139}Institute of Space Sciences (ISS), Bucharest, Romania
\item \Idef{org27399}Institut f\"{u}r Informatik, Johann Wolfgang Goethe-Universit\"{a}t Frankfurt, Frankfurt, Germany
\item \Idef{org1185}Institut f\"{u}r Kernphysik, Johann Wolfgang Goethe-Universit\"{a}t Frankfurt, Frankfurt, Germany
\item \Idef{org1177}Institut f\"{u}r Kernphysik, Technische Universit\"{a}t Darmstadt, Darmstadt, Germany
\item \Idef{org1256}Institut f\"{u}r Kernphysik, Westf\"{a}lische Wilhelms-Universit\"{a}t M\"{u}nster, M\"{u}nster, Germany
\item \Idef{org1246}Instituto de Ciencias Nucleares, Universidad Nacional Aut\'{o}noma de M\'{e}xico, Mexico City, Mexico
\item \Idef{org1247}Instituto de F\'{\i}sica, Universidad Nacional Aut\'{o}noma de M\'{e}xico, Mexico City, Mexico
\item \Idef{org1308}Institut Pluridisciplinaire Hubert Curien (IPHC), Universit\'{e} de Strasbourg, CNRS-IN2P3, Strasbourg, France
\item \Idef{org1182}Joint Institute for Nuclear Research (JINR), Dubna, Russia
\item \Idef{org1199}Kirchhoff-Institut f\"{u}r Physik, Ruprecht-Karls-Universit\"{a}t Heidelberg, Heidelberg, Germany
\item \Idef{org20954}Korea Institute of Science and Technology Information, Daejeon, South Korea
\item \Idef{org1017642}KTO Karatay University, Konya, Turkey
\item \Idef{org1160}Laboratoire de Physique Corpusculaire (LPC), Clermont Universit\'{e}, Universit\'{e} Blaise Pascal, CNRS--IN2P3, Clermont-Ferrand, France
\item \Idef{org1194}Laboratoire de Physique Subatomique et de Cosmologie (LPSC), Universit\'{e} Joseph Fourier, CNRS-IN2P3, Institut Polytechnique de Grenoble, Grenoble, France
\item \Idef{org1187}Laboratori Nazionali di Frascati, INFN, Frascati, Italy
\item \Idef{org1232}Laboratori Nazionali di Legnaro, INFN, Legnaro, Italy
\item \Idef{org1125}Lawrence Berkeley National Laboratory, Berkeley, California, United States
\item \Idef{org1234}Lawrence Livermore National Laboratory, Livermore, California, United States
\item \Idef{org1251}Moscow Engineering Physics Institute, Moscow, Russia
\item \Idef{org1322}National Centre for Nuclear Studies, Warsaw, Poland
\item \Idef{org1140}National Institute for Physics and Nuclear Engineering, Bucharest, Romania
\item \Idef{org1017626}National Institute of Science Education and Research, Bhubaneswar, India
\item \Idef{org1165}Niels Bohr Institute, University of Copenhagen, Copenhagen, Denmark
\item \Idef{org1109}Nikhef, National Institute for Subatomic Physics, Amsterdam, Netherlands
\item \Idef{org1283}Nuclear Physics Institute, Academy of Sciences of the Czech Republic, \v{R}e\v{z} u Prahy, Czech Republic
\item \Idef{org1264}Oak Ridge National Laboratory, Oak Ridge, Tennessee, United States
\item \Idef{org1189}Petersburg Nuclear Physics Institute, Gatchina, Russia
\item \Idef{org1170}Physics Department, Creighton University, Omaha, Nebraska, United States
\item \Idef{org1157}Physics Department, Panjab University, Chandigarh, India
\item \Idef{org1112}Physics Department, University of Athens, Athens, Greece
\item \Idef{org1152}Physics Department, University of Cape Town and  iThemba LABS, National Research Foundation, Somerset West, South Africa
\item \Idef{org1209}Physics Department, University of Jammu, Jammu, India
\item \Idef{org1207}Physics Department, University of Rajasthan, Jaipur, India
\item \Idef{org1200}Physikalisches Institut, Ruprecht-Karls-Universit\"{a}t Heidelberg, Heidelberg, Germany
\item \Idef{org1017688}Politecnico di Torino, Turin, Italy
\item \Idef{org1325}Purdue University, West Lafayette, Indiana, United States
\item \Idef{org1281}Pusan National University, Pusan, South Korea
\item \Idef{org1176}Research Division and ExtreMe Matter Institute EMMI, GSI Helmholtzzentrum f\"ur Schwerionenforschung, Darmstadt, Germany
\item \Idef{org1334}Rudjer Bo\v{s}kovi\'{c} Institute, Zagreb, Croatia
\item \Idef{org1298}Russian Federal Nuclear Center (VNIIEF), Sarov, Russia
\item \Idef{org1252}Russian Research Centre Kurchatov Institute, Moscow, Russia
\item \Idef{org1224}Saha Institute of Nuclear Physics, Kolkata, India
\item \Idef{org1130}School of Physics and Astronomy, University of Birmingham, Birmingham, United Kingdom
\item \Idef{org1338}Secci\'{o}n F\'{\i}sica, Departamento de Ciencias, Pontificia Universidad Cat\'{o}lica del Per\'{u}, Lima, Peru
\item \Idef{org1133}Sezione INFN, Bologna, Italy
\item \Idef{org1271}Sezione INFN, Padova, Italy
\item \Idef{org1286}Sezione INFN, Rome, Italy
\item \Idef{org1146}Sezione INFN, Cagliari, Italy
\item \Idef{org1313}Sezione INFN, Turin, Italy
\item \Idef{org1316}Sezione INFN, Trieste, Italy
\item \Idef{org1115}Sezione INFN, Bari, Italy
\item \Idef{org1155}Sezione INFN, Catania, Italy
\item \Idef{org36377}Nuclear Physics Group, STFC Daresbury Laboratory, Daresbury, United Kingdom
\item \Idef{org1258}SUBATECH, Ecole des Mines de Nantes, Universit\'{e} de Nantes, CNRS-IN2P3, Nantes, France
\item \Idef{org35706}Suranaree University of Technology, Nakhon Ratchasima, Thailand
\item \Idef{org1304}Technical University of Split FESB, Split, Croatia
\item \Idef{org1168}The Henryk Niewodniczanski Institute of Nuclear Physics, Polish Academy of Sciences, Cracow, Poland
\item \Idef{org17361}The University of Texas at Austin, Physics Department, Austin, TX, United States
\item \Idef{org1173}Universidad Aut\'{o}noma de Sinaloa, Culiac\'{a}n, Mexico
\item \Idef{org1296}Universidade de S\~{a}o Paulo (USP), S\~{a}o Paulo, Brazil
\item \Idef{org1149}Universidade Estadual de Campinas (UNICAMP), Campinas, Brazil
\item \Idef{org1239}Universit\'{e} de Lyon, Universit\'{e} Lyon 1, CNRS/IN2P3, IPN-Lyon, Villeurbanne, France
\item \Idef{org1205}University of Houston, Houston, Texas, United States
\item \Idef{org20371}University of Technology and Austrian Academy of Sciences, Vienna, Austria
\item \Idef{org1222}University of Tennessee, Knoxville, Tennessee, United States
\item \Idef{org1310}University of Tokyo, Tokyo, Japan
\item \Idef{org1318}University of Tsukuba, Tsukuba, Japan
\item \Idef{org21360}Eberhard Karls Universit\"{a}t T\"{u}bingen, T\"{u}bingen, Germany
\item \Idef{org1225}Variable Energy Cyclotron Centre, Kolkata, India
\item \Idef{org1306}V.~Fock Institute for Physics, St. Petersburg State University, St. Petersburg, Russia
\item \Idef{org1323}Warsaw University of Technology, Warsaw, Poland
\item \Idef{org1179}Wayne State University, Detroit, Michigan, United States
\item \Idef{org1143}Wigner Research Centre for Physics, Hungarian Academy of Sciences, Budapest, Hungary
\item \Idef{org1260}Yale University, New Haven, Connecticut, United States
\item \Idef{org15649}Yildiz Technical University, Istanbul, Turkey
\item \Idef{org1301}Yonsei University, Seoul, South Korea
\item \Idef{org1327}Zentrum f\"{u}r Technologietransfer und Telekommunikation (ZTT), Fachhochschule Worms, Worms, Germany
\end{Authlist}
\endgroup

\else
\ifbibtex
\bibliographystyle{\bibstname}
\bibliography{biblio}{}
\else
 
\fi
\appendix
\section{Tables}
\label{sec:tables}

\fi
\else
\iffull
\vspace{0.5cm}
\input{refpaper.tex}
\newpage
\appendix
\section{Tables}
\label{sec:tables}

\else
\appendix
\section{Tables}
\label{sec:tables}

\ifbibtex
\bibliographystyle{\bibstname}
\bibliography{biblio}{}
\else
\input{refpaper.tex}
\fi
\fi
\fi
\end{document}